\documentclass[12pt]{iopart}

\input{definitions}

\let\Sectionmark\sectionmark
\def\sectionmark#1{\def\Sectionname{\uppercase{#1}}\Sectionmark{#1}}
\let\Subsectionmark\subsectionmark
\def\subsectionmark#1{\def\Subsectionname{#1}\Subsectionmark{#1}}

\begin{document}
\leftline{Dated: \today}

\title[]{The Virgo O3 run and the impact of the environment}

\author{%
F~Acernese$^{1,2}$, 
M~Agathos$^{3}$, 
A~Ain$^{4}$, 
S~Albanesi$^{5,6}$, 
A~Allocca$^{7,2}$, 
A~Amato$^{8}$, 
T~Andrade$^{9}$, 
N~Andres$^{10}$, 
M~Andr\'es-Carcasona$^{11}$, %
T~Andri\'c$^{12}$, 
S~Ansoldi$^{13,14}$
% ...
}%
\medskip
\address {${}{\ast}$Deceased, April 2021.}%
\medskip
\address {$^{1}$Dipartimento di Farmacia, Universit\`a di Salerno, I-84084 Fisciano, Salerno, Italy }
\address {$^{2}$INFN, Sezione di Napoli, Complesso Universitario di Monte S. Angelo, I-80126 Napoli, Italy }
\address {$^{3}$Theoretisch-Physikalisches Institut, Friedrich-Schiller-Universit\"at Jena, D-07743 Jena, Germany }
\address {$^{4}$INFN, Sezione di Pisa, I-56127 Pisa, Italy }
\address {$^{5}$Dipartimento di Fisica, Universit\`a degli Studi di Torino, I-10125 Torino, Italy }
\address {$^{6}$INFN Sezione di Torino, I-10125 Torino, Italy }
\address {$^{7}$Universit\`a di Napoli ``Federico II'', Complesso Universitario di Monte S. Angelo, I-80126 Napoli, Italy }
\address {$^{8}$Universit\'e de Lyon, Universit\'e Claude Bernard Lyon 1, CNRS, Institut Lumi\`ere Mati\`ere, F-69622 Villeurbanne, France }
\address {$^{9}$Institut de Ci\`encies del Cosmos (ICCUB), Universitat de Barcelona, C/ Mart\'{\i } i Franqu\`es 1, Barcelona, 08028, Spain }
\address {$^{10}$Univ. Savoie Mont Blanc, CNRS, Laboratoire d'Annecy de Physique des Particules - IN2P3, F-74000 Annecy, France }
\address {$^{11}$Institut de F\'{\i }sica d'Altes Energies (IFAE), Barcelona Institute of Science and Technology, and ICREA, E-08193 Barcelona, Spain }
\address {$^{12}$Gran Sasso Science Institute (GSSI), I-67100 L'Aquila, Italy }
\address {$^{13}$Dipartimento di Scienze Matematiche, Informatiche e Fisiche, Universit\`a di Udine, I-33100 Udine, Italy }
\address {$^{14}$INFN, Sezione di Trieste, I-34127 Trieste, Italy }
\address {$^{15}$Artemis, Universit\'e C\^ote d'Azur, Observatoire de la C\^ote d'Azur, CNRS, F-06304 Nice, France }
\address {$^{16}$GRAPPA, Anton Pannekoek Institute for Astronomy and Institute for High-Energy Physics, University of Amsterdam, Science Park 904, 1098 XH Amsterdam, Netherlands }
\address {$^{17}$National and Kapodistrian University of Athens, School of Science Building, 2nd floor, Panepistimiopolis, 15771 Ilissia, Greece }
\address {$^{18}$INFN, Sezione di Perugia, I-06123 Perugia, Italy }
\address {$^{19}$Universit\`a di Camerino, Dipartimento di Fisica, I-62032 Camerino, Italy }
\address {$^{20}$Universit\'e de Paris, CNRS, Astroparticule et Cosmologie, F-75006 Paris, France }
\address {$^{21}$Universit\'e Paris-Saclay, CNRS/IN2P3, IJCLab, 91405 Orsay, France }
\address {$^{22}$European Gravitational Observatory (EGO), I-56021 Cascina, Pisa, Italy }
\address {$^{23}$Universit\`a degli Studi di Urbino ``Carlo Bo'', I-61029 Urbino, Italy }
\address {$^{24}$INFN, Sezione di Firenze, I-50019 Sesto Fiorentino, Firenze, Italy }
\address {$^{25}$INFN, Sezione di Roma, I-00185 Roma, Italy }
\address {$^{26}$Universit\'e Lyon, Universit\'e Claude Bernard Lyon 1, CNRS, IP2I Lyon / IN2P3, UMR 5822, F-69622 Villeurbanne, France }
\address {$^{27}$Universit\'e catholique de Louvain, B-1348 Louvain-la-Neuve, Belgium }
\address {$^{28}$Nikhef, Science Park 105, 1098 XG Amsterdam, Netherlands }
\address {$^{29}$Institute for Gravitational and Subatomic Physics (GRASP), Utrecht University, Princetonplein 1, 3584 CC Utrecht, Netherlands }
\address {$^{30}$Universit\'e de Li\`ege, B-4000 Li\`ege, Belgium }
\address {$^{31}$Universit\`a degli Studi di Milano-Bicocca, I-20126 Milano, Italy }
\address {$^{32}$INFN, Sezione di Milano-Bicocca, I-20126 Milano, Italy }
\address {$^{33}$INAF, Osservatorio Astronomico di Brera sede di Merate, I-23807 Merate, Lecco, Italy }
\address {$^{34}$Dipartimento di Medicina, Chirurgia e Odontoiatria ``Scuola Medica Salernitana'', Universit\`a di Salerno, I-84081 Baronissi, Salerno, Italy }
\address {$^{35}$Wigner RCP, RMKI, H-1121 Budapest, Konkoly Thege Mikl\'os \'ut 29-33, Hungary }
\address {$^{36}$Universit\`a di Pisa, I-56127 Pisa, Italy }
\address {$^{37}$Universit\`a di Perugia, I-06123 Perugia, Italy }
\address {$^{38}$Universit\`a di Padova, Dipartimento di Fisica e Astronomia, I-35131 Padova, Italy }
\address {$^{39}$INFN, Sezione di Padova, I-35131 Padova, Italy }
\address {$^{40}$Universiteit Gent, B-9000 Gent, Belgium }
\address {$^{41}$Nicolaus Copernicus Astronomical Center, Polish Academy of Sciences, 00-716, Warsaw, Poland }
\address {$^{42}$Dipartimento di Ingegneria, Universit\`a del Sannio, I-82100 Benevento, Italy }
\address {$^{43}$Departamento de Matem\'aticas, Universitat Aut\`onoma de Barcelona, Edificio C Facultad de Ciencias 08193 Bellaterra (Barcelona), Spain }
\address {$^{44}$INFN, Sezione di Genova, I-16146 Genova, Italy }
\address {$^{45}$Vrije Universiteit Amsterdam, 1081 HV Amsterdam, Netherlands }
\address {$^{46}$Universit\`a di Trento, Dipartimento di Fisica, I-38123 Povo, Trento, Italy }
\address {$^{47}$INFN, Trento Institute for Fundamental Physics and Applications, I-38123 Povo, Trento, Italy }
\address {$^{48}$Dipartimento di Fisica ``E.R. Caianiello'', Universit\`a di Salerno, I-84084 Fisciano, Salerno, Italy }
\address {$^{49}$INFN, Sezione di Napoli, Gruppo Collegato di Salerno, Complesso Universitario di Monte S. Angelo, I-80126 Napoli, Italy }
\address {$^{50}$Universit\`a di Roma ``La Sapienza'', I-00185 Roma, Italy }
\address {$^{51}$Univ Rennes, CNRS, Institut FOTON - UMR6082, F-3500 Rennes, France }
\address {$^{52}$INFN, Laboratori Nazionali del Gran Sasso, I-67100 Assergi, Italy }
\address {$^{53}$Laboratoire Kastler Brossel, Sorbonne Universit\'e, CNRS, ENS-Universit\'e PSL, Coll\`ege de France, F-75005 Paris, France }
\address {$^{54}$Astronomical Observatory Warsaw University, 00-478 Warsaw, Poland }
\address {$^{55}$L2IT, Laboratoire des 2 Infinis - Toulouse, Universit\'e de Toulouse, CNRS/IN2P3, UPS, F-31062 Toulouse Cedex 9, France }
\address {$^{56}$Dipartimento di Fisica, Universit\`a degli Studi di Genova, I-16146 Genova, Italy }
\address {$^{57}$Universit\`a di Bologna, Dipartimento di Fisica e Astronomia, I-40127 Bologna, Italy }
\address {$^{58}$Universit\`a degli Studi di Sassari, I-07100 Sassari, Italy }
\address {$^{59}$INFN, Laboratori Nazionali del Sud, I-95125 Catania, Italy }
\address {$^{60}$Universit\`a di Roma Tor Vergata, I-00133 Roma, Italy }
\address {$^{61}$INFN, Sezione di Roma Tor Vergata, I-00133 Roma, Italy }
\address {$^{62}$Departamento de Astronom\'{\i }a y Astrof\'{\i }sica, Universitat de Val\`encia, E-46100 Burjassot, Val\`encia, Spain }
\address {$^{63}$Dipartimento di Ingegneria Industriale (DIIN), Universit\`a di Salerno, I-84084 Fisciano, Salerno, Italy }
\address {$^{64}$INAF, Osservatorio Astronomico di Padova, I-35122 Padova, Italy }
\address {$^{65}$Universit\'e libre de Bruxelles, Avenue Franklin Roosevelt 50 - 1050 Bruxelles, Belgium }
\address {$^{66}$Departamento de Matem\'aticas, Universitat de Val\`encia, E-46100 Burjassot, Val\`encia, Spain }
\address {$^{67}$University of Minnesota, Minneapolis, MN 55455, USA }
\address {$^{68}$Scuola Normale Superiore, Piazza dei Cavalieri, 7 - 56126 Pisa, Italy }
\address {$^{69}$Maastricht University, P.O. Box 616, 6200 MD Maastricht, Netherlands }
\address {$^{70}$Universit\'e Lyon, Universit\'e Claude Bernard Lyon 1, CNRS, Laboratoire des Mat\'eriaux Avanc\'es (LMA), IP2I Lyon / IN2P3, UMR 5822, F-69622 Villeurbanne, France }
\address {$^{71}$Dipartimento di Scienze Matematiche, Fisiche e Informatiche, Universit\`a di Parma, I-43124 Parma, Italy }
\address {$^{72}$INFN, Sezione di Milano Bicocca, Gruppo Collegato di Parma, I-43124 Parma, Italy }
\address {$^{73}$Universit\'e de Strasbourg, CNRS, IPHC UMR 7178, F-67000 Strasbourg, France }
\address {$^{74}$Institute for Nuclear Research, Bem t'er 18/c, H-4026 Debrecen, Hungary }
\address {$^{75}$CNR-SPIN, c/o Universit\`a di Salerno, I-84084 Fisciano, Salerno, Italy }
\address {$^{76}$Scuola di Ingegneria, Universit\`a della Basilicata, I-85100 Potenza, Italy }
\address {$^{77}$Gravitational Wave Science Project, National Astronomical Observatory of Japan (NAOJ), Mitaka City, Tokyo 181-8588, Japan }
\address {$^{78}$Observatori Astron\`omic, Universitat de Val\`encia, E-46980 Paterna, Val\`encia, Spain }
\address {$^{79}$Centro de F\'{\i }sica das Universidades do Minho e do Porto, Universidade do Minho, Campus de Gualtar, PT-4710 - 057 Braga, Portugal }
\address {$^{80}$Istituto Nazionale di Geofisica e Vulcanologia, I-56125 Pisa, Italy }
\address {$^{81}$INAF, Osservatorio Astronomico di Capodimonte, I-80131 Napoli, Italy }
\address {$^{82}$Aristotle University of Thessaloniki, University Campus, 54124 Thessaloniki, Greece }
\address {$^{83}$Universiteit Antwerpen, Prinsstraat 13, 2000 Antwerpen, Belgium }
\address {$^{84}$University of Bia{\l }ystok, 15-424 Bia{\l }ystok, Poland }
\address {$^{85}$Institute for High-Energy Physics, University of Amsterdam, Science Park 904, 1098 XH Amsterdam, Netherlands }
\address {$^{86}$Institute of Mathematics, Polish Academy of Sciences, 00656 Warsaw, Poland }
\address {$^{87}$National Center for Nuclear Research, 05-400 {\' S}wierk-Otwock, Poland }
\address {$^{88}$Laboratoire Lagrange, Universit\'e C\^ote d'Azur, Observatoire C\^ote d'Azur, CNRS, F-06304 Nice, France }
\address {$^{89}$NAVIER, \'{E}cole des Ponts, Univ Gustave Eiffel, CNRS, Marne-la-Vall\'{e}e, France }
\address {$^{90}$Universit\`a di Firenze, Sesto Fiorentino I-50019, Italy }
\address {$^{91}$INFN, Sezione di Roma Tre, I-00146 Roma, Italy }
\address {$^{92}$ESPCI, CNRS, F-75005 Paris, France }
\address {$^{93}$School of Physics Science and Engineering, Tongji University, Shanghai 200092, China }
\address {$^{94}$Dipartimento di Fisica, Universit\`a di Trieste, I-34127 Trieste, Italy }
\address {$^{95}$Max Planck Institute for Gravitational Physics (Albert Einstein Institute), D-30167 Hannover, Germany }
\address {$^{96}$Leibniz Universit\"at Hannover, D-30167 Hannover, Germany }
\address {$^{97}$Institut des Hautes Etudes Scientifiques, F-91440 Bures-sur-Yvette, France }
\address {$^{98}$Istituto Nazionale di Geofisica e Vulcanologia, I-40100 Bologna, Italy }
\address {$^{99}$Consiglio Nazionale delle Ricerche - Istituto dei Sistemi Complessi, Piazzale Aldo Moro 5, I-00185 Roma, Italy }
\address {$^{100}$Museo Storico della Fisica e Centro Studi e Ricerche ``Enrico Fermi'', I-00184 Roma, Italy }
\address {$^{101}$Dipartimento di Matematica e Fisica, Universit\`a degli Studi Roma Tre, I-00146 Roma, Italy }
\address {$^{102}$Universit\`a di Trento, Dipartimento di Matematica, I-38123 Povo, Trento, Italy }
\address {$^{103}$Departamento de Matem\'atica da Universidade de Aveiro and Centre for Research and Development in Mathematics and Applications, Campus de Santiago, 3810-183 Aveiro, Portugal }
\address {$^{104}$INAF, Osservatorio Astrofisico di Arcetri, Largo E. Fermi 5, I-50125 Firenze, Italy }
\address {$^{105}$Istituto di Astrofisica e Planetologia Spaziali di Roma, Via del Fosso del Cavaliere, 100, 00133 Roma RM, Italy }
\address {$^{106}$LIGO Livingston Observatory, Livingston, LA 70754, USA }
\address {$^{107}$Dipartimento di Scienze Aziendali - Management and Innovation Systems (DISA-MIS), Universit\`a di Salerno, I-84084 Fisciano, Salerno, Italy }
\address {$^{108}$Van Swinderen Institute for Particle Physics and Gravity, University of Groningen, Nijenborgh 4, 9747 AG Groningen, Netherlands }
\address {$^{109}$Vrije Universiteit Brussel, Pleinlaan 2, 1050 Brussel, Belgium }
\address {$^{110}$Eindhoven University of Technology, Postbus 513, 5600 MB Eindhoven, Netherlands }

\begin{abstract}
Sources of geophysical noise (such as wind, sea waves and earthquakes) or of anthropogenic noise impact ground-based gravitational-wave interferometric detectors, causing transient sensitivity worsening and gaps in data taking.
During the one year-long third Observing Run (O3: from April 01, 2019 to March 27, 2020), the Virgo Collaboration collected a statistically significant dataset, used in this article to study the response of the detector to a variety of environmental conditions. We correlated environmental parameters to global detector performance, such as observation range, duty cycle and control losses. Where possible, we identified weaknesses in the detector that will be used to 
elaborate strategies in order to improve Virgo robustness against external disturbances for the next data taking period, O4, currently planned to start at the end of 2022. 
The lessons learned could also provide useful insights for the design of the next generation of ground-based interferometers.
\end{abstract}

\maketitle

\tableofcontents

\clearpage

\setlength{\parindent}{0pt}
\setlength{\parskip}{\medskipamount}

\section{Introduction}
\label{section:introduction}
\markboth{\thesection. \Sectionname}{}
The past decade has seen the ramp-up of the second-generation ('Advanced') earth-based gravitational-wave (GW) detectors. Design improvements and technological upgrades have paved the way to the first direct detections of GWs by the global network made up of the two aLIGO instruments~\cite{TheLIGOScientific:2014jea} (located in the USA: Hanford, WA and Livingston, LA) and of the Advanced Virgo detector~\cite{TheVirgo:2014hva} (located in Cascina, Italy). The main results achieved by the LIGO Scientific Collaboration and the Virgo Collaboration~-- recently joined by the KAGRA collaboration whose detector~\cite{10.1093/ptep/ptab018} (located in Kamioka, Japan, under the Ikenoue mountain) is nearing completion~-- include the first detection of a binary black hole merger (GW150914~\cite{Abbott:2016blz}); the first detection of a binary neutron star (BNS) merger (GW170817~\cite{TheLIGOScientific:2017qsa}) that lead to the birth of multi-messenger astronomy with GW~\cite{GBM:2017lvd}; and now dozens of detections of compact binary mergers that add up in a GW Transient Catalogue regularly updated~\cite{LIGOScientific:2018mvr,Abbott:2020niy,GWTC3}. These detections contribute to opening a new window onto the Universe by providing insights to the populations of compact objects and the binary merger rates~\cite{Abbott_2021}; they also allow scientists to perform stringent tests of general relativity~\cite{PhysRevD.103.122002} in a new regime of gravitation never probed before.

The operation of ground-based GW detectors is organized into successive steps forming a recurring sequence over the years: upgrades; commissioning and sensitivity improvement (the so-called {\em noise hunting} phase); data-taking periods called observing runs (or simply runs and labelled O$n$). So far there have been three runs for the global network of advanced detectors.
\begin{itemize}
\item O1 (09/2015~-- 01/2016) with only the two LIGO detectors taking data;
\item O2 (11/2016~-- 08/2017) with Virgo joining LIGO on August 01, 2017;
\item finally O3 (04/2019~-- 03/2020), that saw the three detectors take data jointly during 11 months in total: 6 months first (a run called O3a), followed by a 1-month break (October 2019) and then another period of 5 months of data taking (O3b), interrupted about a month earlier than expected due to the worldwide COVID-19 pandemic.
\end{itemize}

The above listing shows that the O3 run was the first {\em long} data-taking period for the Advanced Virgo detector. Therefore, we have used the wealth of unprecedented data collected during this year to make an in-depth analysis of the instrument performance. In this article, we study the impact of the environment on Advanced Virgo, along the lines of previous publications from Virgo~\cite{Virgo_env_O3}, LIGO~\cite{LIGO_env_O3} or KAGRA~\cite{Washimi_2021}. We focus on various types of seismic noises, on earthquakes and on bad weather periods. We also briefly investigate the effect of other possible disturbances: magnetic noise, lightning and cosmic muons. Our goal is threefold: to quantify how the Virgo sensitivity and duty cycle depend on these external parameters; to use this knowledge to prepare the next run, O4, scheduled to start in the second semester 2022; finally, to build experience for future GW detectors, in particular for the Einstein Telescope project~\cite{Punturo_2010}.

The Virgo detector is located in Italy at EGO, the European Gravitational Observatory, in the municipality of Cascina. The EGO site is in the countryside, about 12~km south-east of Pisa and about 17~km east of the Tyrrhenian coast. Virgo is not far from some industrial and commercial sites that can generate noise. Within 7~km from EGO there are: elevated highways, railway tracks, wind turbines, earth quarries, high-voltage power lines (electricity pylons and overhead line segments) and the Pisa airport. To avoid pressure waves potentially shaking the ground, a no-fly zone has been enforced in a cylindrical volume (600~m radius and height) above each of the Virgo experimental buildings.

Advanced Virgo is a power-recycled Michelson interferometer with Fabry-Perot cavities in its 3~km-long arms. All core optics are suspended to long suspensions, called the superattenuators~\cite{ACERNESE2004629}, that have a twofold use: first, to isolate as much as possible the mirrors from seismic motions (both vertical and longitudinal), and then to control very accurately their positions in all six degrees of freedom. Many feedback systems are used to bring the detector to its working point and maintain it there~\cite{ACERNESE2020102386,galaxies8040085}. This state~-- the same for O2 and O3: the Michelson interferometer on a dark fringe, the Fabry-Perot and power recycling cavities in resonance~-- is the only one in which the detector is sensitive to the passing of GWs.

During a run, priority is obviously given to taking data of quality good enough to be included in physics analysis. In that case, Virgo is said to be in {\em Science mode}. During O3, the average duty cycle in Science mode has been around 76\%~\cite{o3virgodetchar}, with the remaining time almost equally divided into three categories.
\begin{itemize}
\item Control acquisition and adjustment phases, to restore the working point and restart taking data in Science mode;
\item Recurring controlled actions on the detector: maintenance (usually a few hours on Tuesday mornings local time), calibration (usually every Wednesday evening) or commissioning (measurements, working point tuning or tailored improvements: sessions organized when the need arises);
\item Problems preventing a smooth running of the detector.
\end{itemize}

The article is organized as follows. Section~\ref{section:virgo} describes the environmental monitoring of the Virgo detector during the O3 run. Section~\ref{section:seismic_noise} is dedicated to the different seismic noise contributions (either natural or human-related): how to disentangle them, how to monitor them and what their impacts on the detector are in terms of sensitivity and duty cycle. Section~\ref{section:earthquakes} provides an analysis of the impact of earthquakes on the detector.
Section~\ref{section:bad_weather} studies the impact of bad weather on data quality and duty cycle, disentangling contributions from sea activity and wind. Section~\ref{section:other} goes through other environment impacts: magnetic noise, lightning and a study of the cosmic muon rate on the Virgo central building. Then, Section~\ref{section:outlook} concludes this article by opening outlooks to the future O4 run. Finally, \ref{section:lock_losses} provides a detailed and quite complete classification of the control losses during the O3 run. Although that study has a scope broader than the present article, it is included here for reference and also because its results were used, in particular to find out which control losses were due to earthquakes.

\section{The Virgo environmental monitoring during O3}
\label{section:virgo}
\markboth{\thesection. \Sectionname}{}
The Virgo detector is equipped with a large set of probes used to monitor the conditions of the surrounding environment. Since these conditions can influence the detector response, or even mimic a GW event, it is very important to track their evolution, to assess the right working condition of the detector or to use them as veto against possible fake signals.

\begin{figure}[!htbp]
	\centering
	\includegraphics[width=0.9\textwidth]{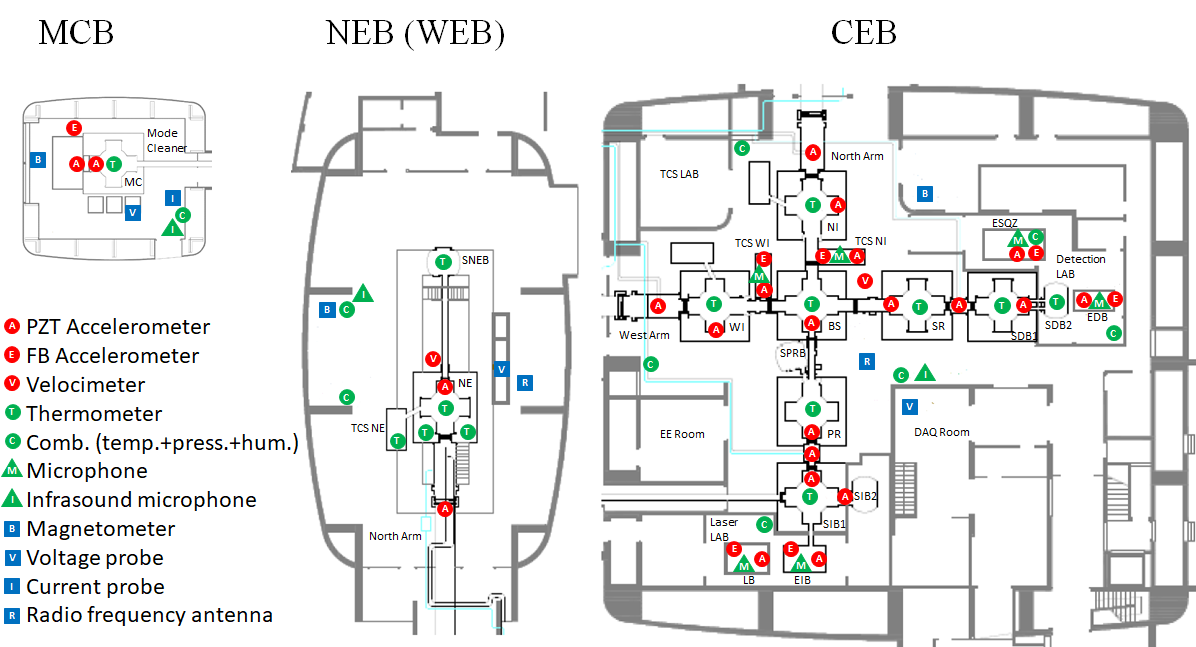}
	\caption{Location of the probes used for the Virgo environmental monitoring system. Maps of most relevant building are shown: left MCB, middle NEB, right CEB. The WEB is very similar to NEB and is not shown.}
	\label{fig:env_map}
\end{figure}

The set of probes and their conditioning electronics constitute the Environmental Monitoring System (EMS). The  EMS was initially composed by a few tens of environmental probes (EPs)~\cite{env_mon1} and then improved during the detector upgrades that occurred in the past years. During O3, the total number of channels belonging to EMS was about 420.

The EMS is also helpful to understand the origin of some noise sources affecting the detector sensitivity. Indeed it was largely used during the commissioning phase following each detector upgrade, to recover and improve the Virgo performance in terms of sensitivity and duty cycle~\cite{Virgo_env_O3}.

Data acquired for EMS can be grouped in two classes depending on the sample rate used for the different EPs. High-rate, or fast class, includes those EPs acquired at rate up to 20~kHz like seismometers, piezoelectric (PZT) accelerometers, force balance (FB) accelerometer, magnetometers, microphones, voltage and current sensors, radio-frequency (RF) antennas, while low-rate or slow class includes temperature, pressure, humidity, weather and lightning probes, acquired at 1~Hz rate.

\begin{table}[!htbp]
\begin{tabular}{ccc}
\hline
Type	& Model & Frequency Band \\
\hline
Seismometer	& Guralp CMG-40T & 0.01 -- 50~Hz \\
FB Accelerometer & Kinemetrics FBA ES-T & 0.1 -- 200~Hz \\
PZT Accelerometer	& Wilcoxon 731-207 or PCB 393B12 & 1~Hz -- 1~kHz\\
Magnetometer & Metronics MFS-06 or MFS-06e & 0.1 mHz -- 10 kHz \\
Microphone & Br\"uel \& Kj{\ae}r 4190 or 4193 & 0.1 -- 10 kHz \\
RF antenna & AAS STA 5 A/D/0.01-100 & 10 kHz -- 100 MHz \\
Voltage probe & Talema 0015P1-2-009 & DC -- 10 kHz \\
Current probe & IME 0015P1-2-009 & DC -- 10 kHz \\
Temperature probe & Analog Device AD590 & DC -- 0.5~Hz \\
Humidity probe & Honeywell HIH-5031-001 & DC -- 0.5~Hz \\
Pressure probe & NXP MPXA4115A6U & DC -- 0.5~Hz \\
Weather station & Davis Advantage Pro 2 & DC -- 0.3~Hz\\
Lightning detector & Boltek LD 250 & DC -- 0.5~Hz\\
\hline
\end{tabular}
\caption{Characteristics of the Virgo environmental probes used during O3.}
\label{tab:env_tab}
\end{table}

\begin{figure}[!htbp]
	\centering
	\includegraphics[width=0.95\textwidth]{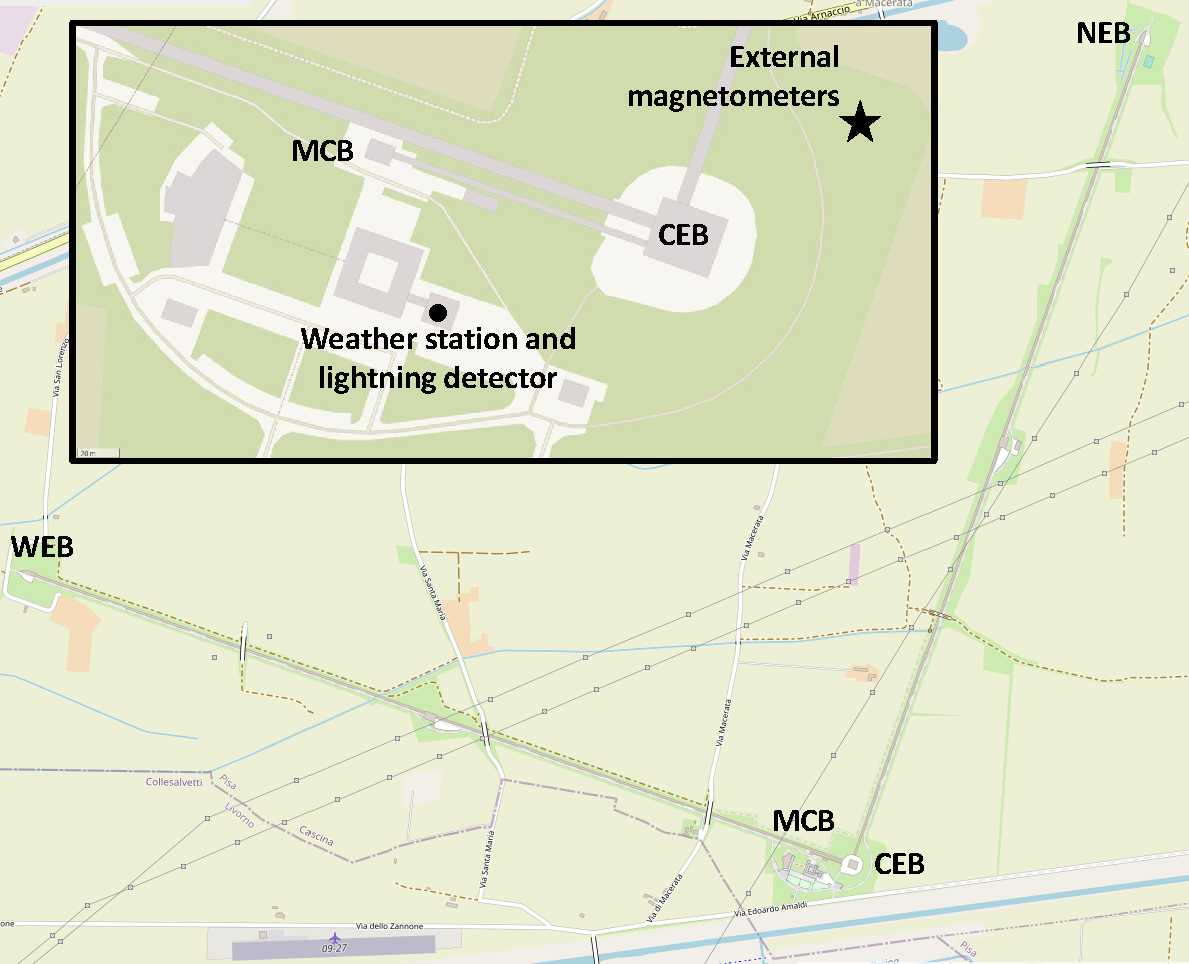}
	\caption{Map~\cite{OpenStreetMap} of the EGO site showing the Virgo detector and in particular the location of the main buildings identified in the text. The central insert shows a zoom around the interferometer vertex, with the CEB and MCB highlighted. The Mode-Cleaner cavity is 144~m-long, while the Virgo arms are 3-km long.}
	\label{fig:virgo_map}
\end{figure}

The main characteristics (type, model and frequency band) of the EPs in use during O3 are listed in Table~\ref{tab:env_tab}. 
Figure~\ref{fig:env_map} shows the arrangement of the EPs inside the main Virgo buildings. Most probes are located in the experimental halls of the relevant buildings of the detector: Central Building (CEB), North and West End Buildings (NEB and WEB) and Mode Cleaner Building (MCB). Usually, the probes are in contact with critical elements of the detector, like the walls of the vacuum chambers containing the test mass suspensions, or the optical benches hosting the laser injection and GW detection systems. Figure~\ref{fig:virgo_map} shows a bird eye's view of the Virgo detector at EGO, with an emphasis on the location of the buildings that are identified in this article.

Few probes are placed outside the buildings, namely the weather station, the lightning detector and two additional magnetometers~-- see Fig.~\ref{fig:virgo_map}. These two low-noise induction coil magnetometers are deployed at 0.5~m depth in the soil, at about 100~m from the CEB, oriented along the geographic North and West directions. Their data are shared in real time with the EM antenna network "Radio waves below 22 kHz"~\cite{vlf_website}.

\section{Seismic noise}
\label{section:seismic_noise}
\markboth{\thesection. \Sectionname}{}
In this section we introduce the main sources of seismic noise at EGO. They are disentangled and monitored by examining seismic probes in specific frequency bands. We provide a statistical description of the noise and evidence its main recurring features. Then, we describe how they impacted on the detector during the O3 run.

\subsection{The seismic frequency bands and their evolution during the O3 run}

\begin{figure}[!htbp]
   \begin{center}
	\includegraphics[width=0.95\textwidth]{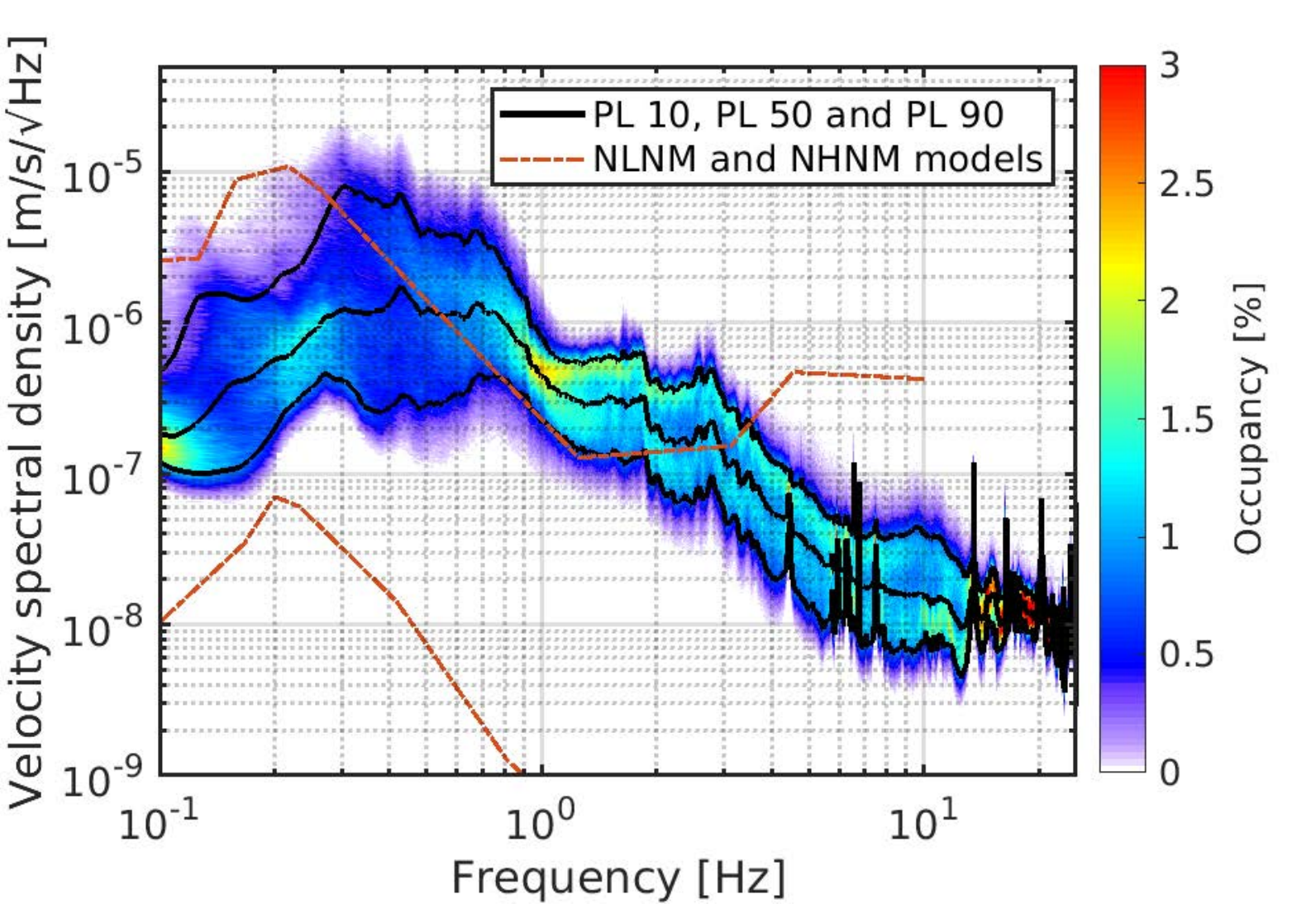}
   \end{center}
	\caption{
	Variability of horizontal velocity of the Virgo NEB ground floor during O3. 
The quantity shown is the 2D histogram of the E-W velocity amplitude spectral density computed for the whole dataset 
recorded during O3 divided into 128~second long chunks. All intereferometer maintenance periods are excluded from the computation.
The color scale indicates the percent occupancy of histogram bins. The superposed continuous curves show different percentile 
levels (labelled {\em PL} on the plot): 10\% (gray), 50\% (black) and 90\% (gray as well). The two red dashed curves 
correspond to the Peterson low-noise ('NLNM') and high-noise ('NHNM') models~\cite{Peterson}.
		}
\label{fig:seismic_percentiles}
\end{figure}

The seismic wavefield at EGO,  the site of the Virgo detector, is the sum of several sources~\cite{Koley}.
Seismic spectrum variability during the O3 run is illustrated in Fig.~\ref{fig:seismic_percentiles}. 
The largest contribution to seismic ground motion in the frequency range between 0.1~Hz and 1~Hz, referred to as {\it microseism}, is due to the interaction between shallow water sea waves and the bottom of the sea~\cite{longuet1950theory,cessaro1994sources}. At EGO, the prevailing microseimic peak is around 0.35~Hz. 

\begin{figure}[!htbp]
\centering
{
\includegraphics[width=0.98\columnwidth]{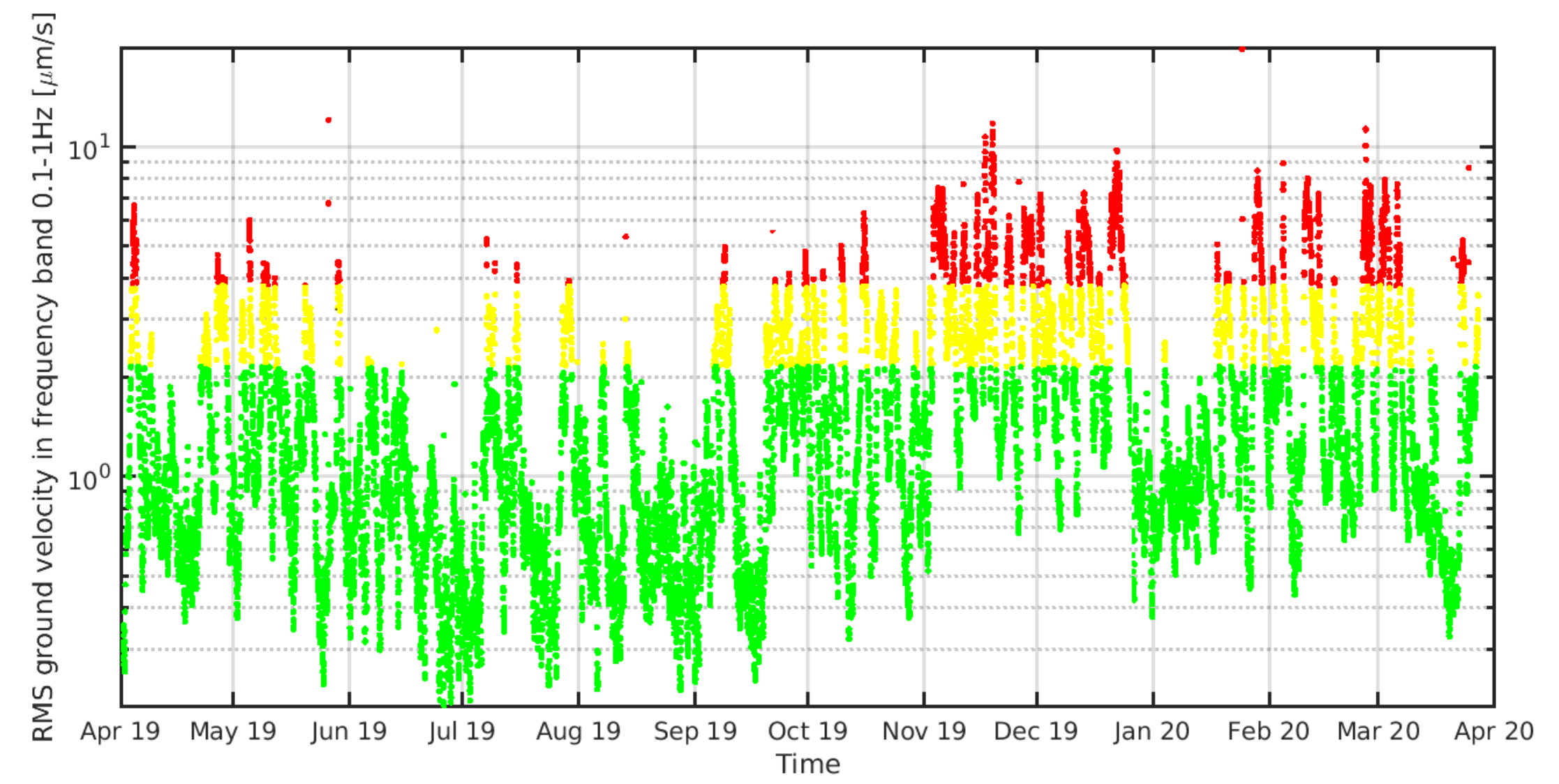} 
}
\caption{Evolution during O3 of seismic RMS in the 0.1 to 1~Hz frequency band. Data colored in yellow and red exceed the 75$^{th}$ and 90$^{th}$ percentile, respectively.} 
\label{fig:sea_activity}
\end{figure} 

\begin{figure}[!htbp]
    \begin{center}
      \includegraphics[width=0.98\columnwidth]{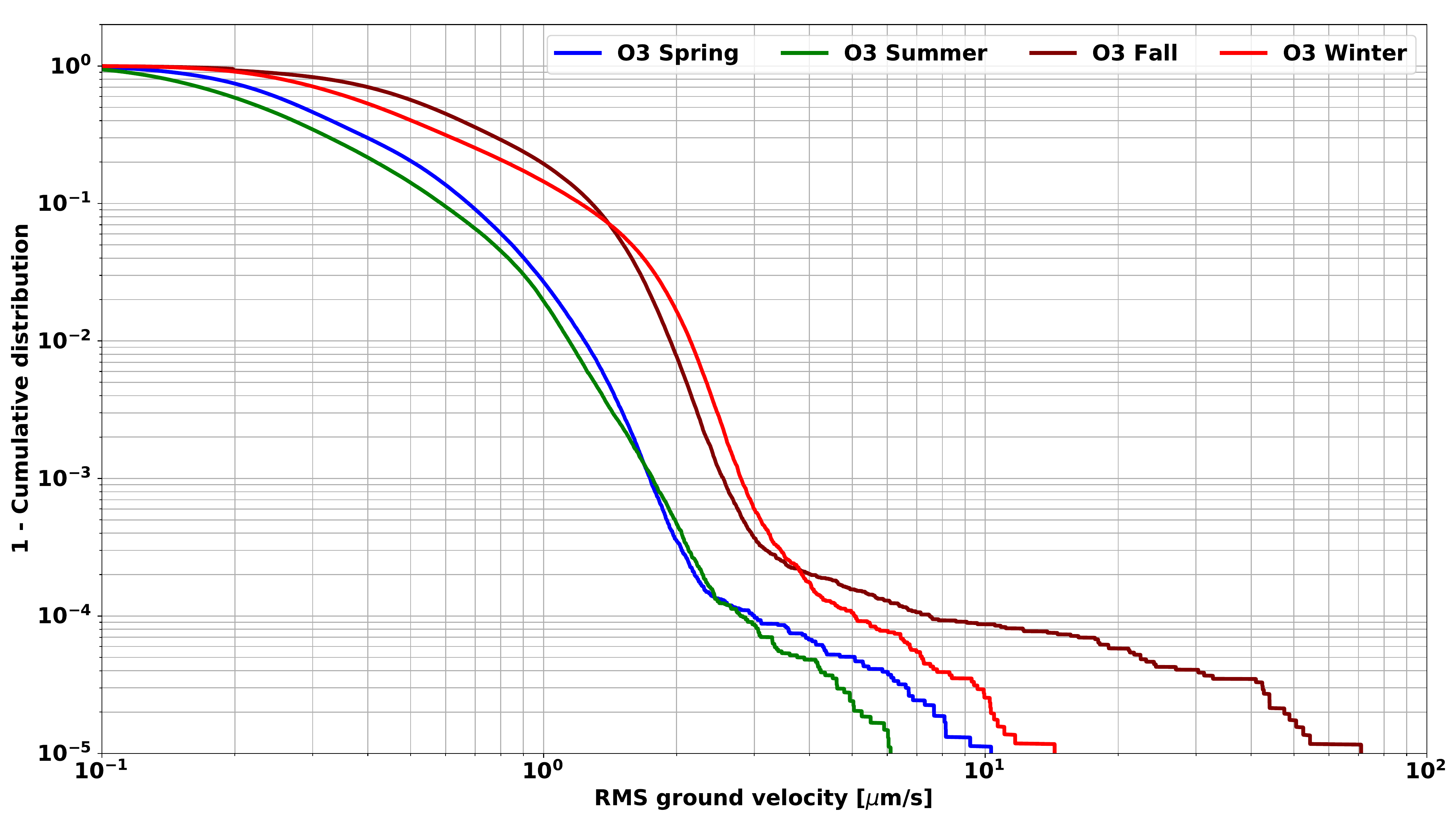}
    \end{center}
    \caption{Cumulative distribution of microseism in the frequency band 0.1-1~Hz (dominated by sea activity), measured at EGO during each season in 2019-2020.}
    \label{fig:microseism_rms}
\end{figure}

Figure~\ref{fig:sea_activity} shows the time evolution of microseism during the O3 run, while Fig.~\ref{fig:microseism_rms} shows the corresponding cumulative distribution, split by season. Microseism intensity follows seasonal variations, being larger in fall and winter, due to the stronger wind and sea activity.

Above 1~Hz, anthropogenic sources dominate the spectrum. 
Heavy vehicles (trucks and alike) on ${\sim} 1$~km distant elevated roads are the prevailing source of seismic noise in the 1-10~Hz band~\cite{Koley}.

\begin{figure}[!htbp]
    \begin{center}
      \includegraphics[width=0.98\columnwidth]{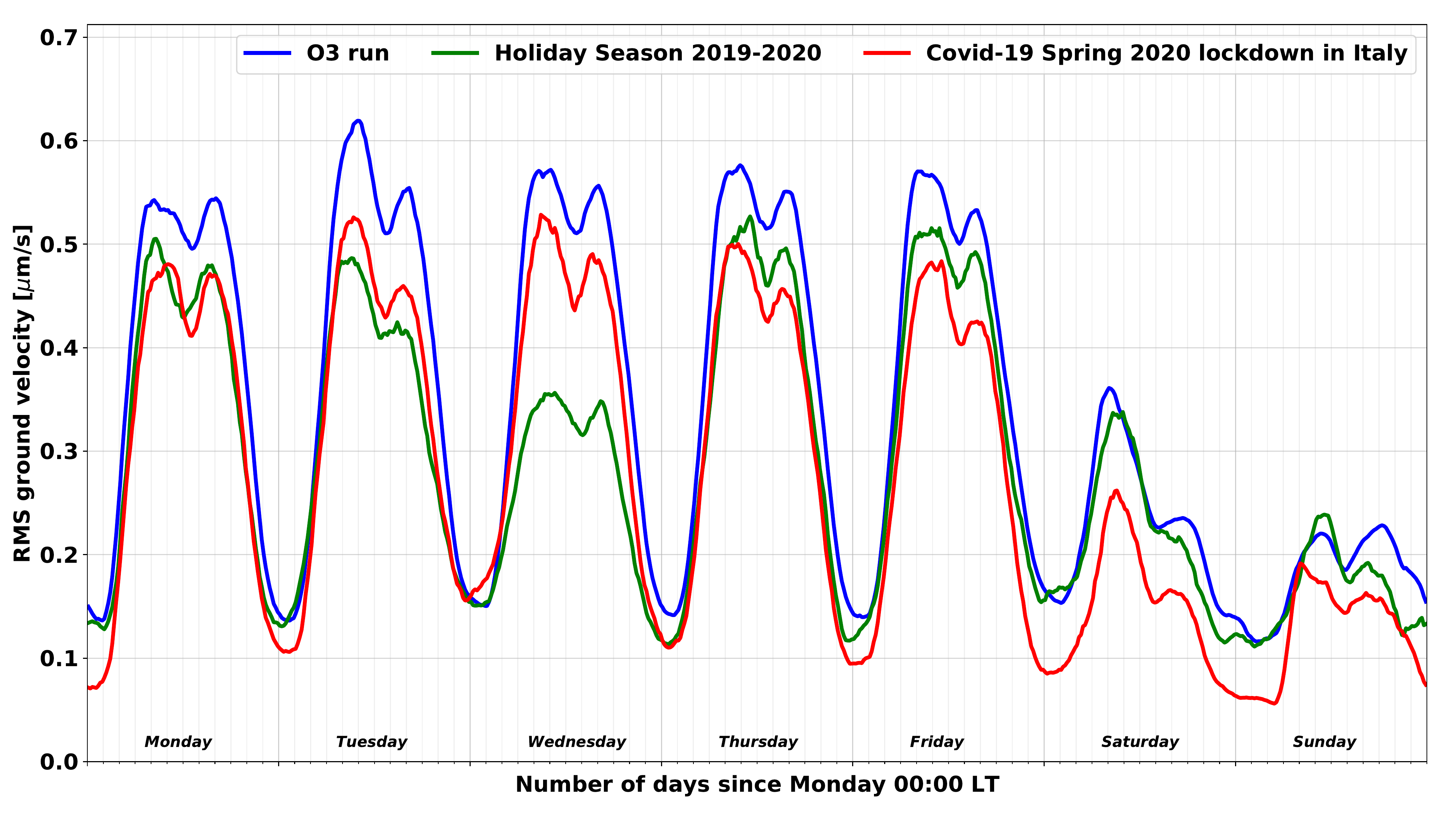}
    \end{center}
    \caption{Average evolution on a weekly basis of the seismic anthropogenic noise (frequency band: 1-5~Hz) measured at EGO during different times in 2019-2020.}
    \label{fig:anthropogenic}
\end{figure}

As illustrated in Fig.~\ref{fig:anthropogenic}, the RMS of seismic noise in the 1-5~Hz band follows a working day/night cycle with higher levels during working hours (from 8:00 to 17:00 local time~-- LT), with small reduction during lunch break (12:00-14:00~LT) and minima during week-ends and holidays. The blue curve, used as reference, covers the whole O3 run. The green curve is based on a 4-week period, from Monday 16 December, 2019 to Sunday 12 January, 2020: the noise reduction during the two consecutive Wednesdays, Christmas 2019 and the New Year's Day 2020, is quite impressive. A significant reduction of the anthropogenic noise is also visible during the Spring 2020 lockdown in Italy, due to the COVID-19 pandemic (red curve, covering a 8-week period from 09 March to 03 May). That decrease is smaller than for the Christmas and New Year holidays but it is more global as it is visible for all days of the week.

\begin{figure}[!htbp]
    \begin{center}
      \includegraphics[width=0.98\columnwidth]{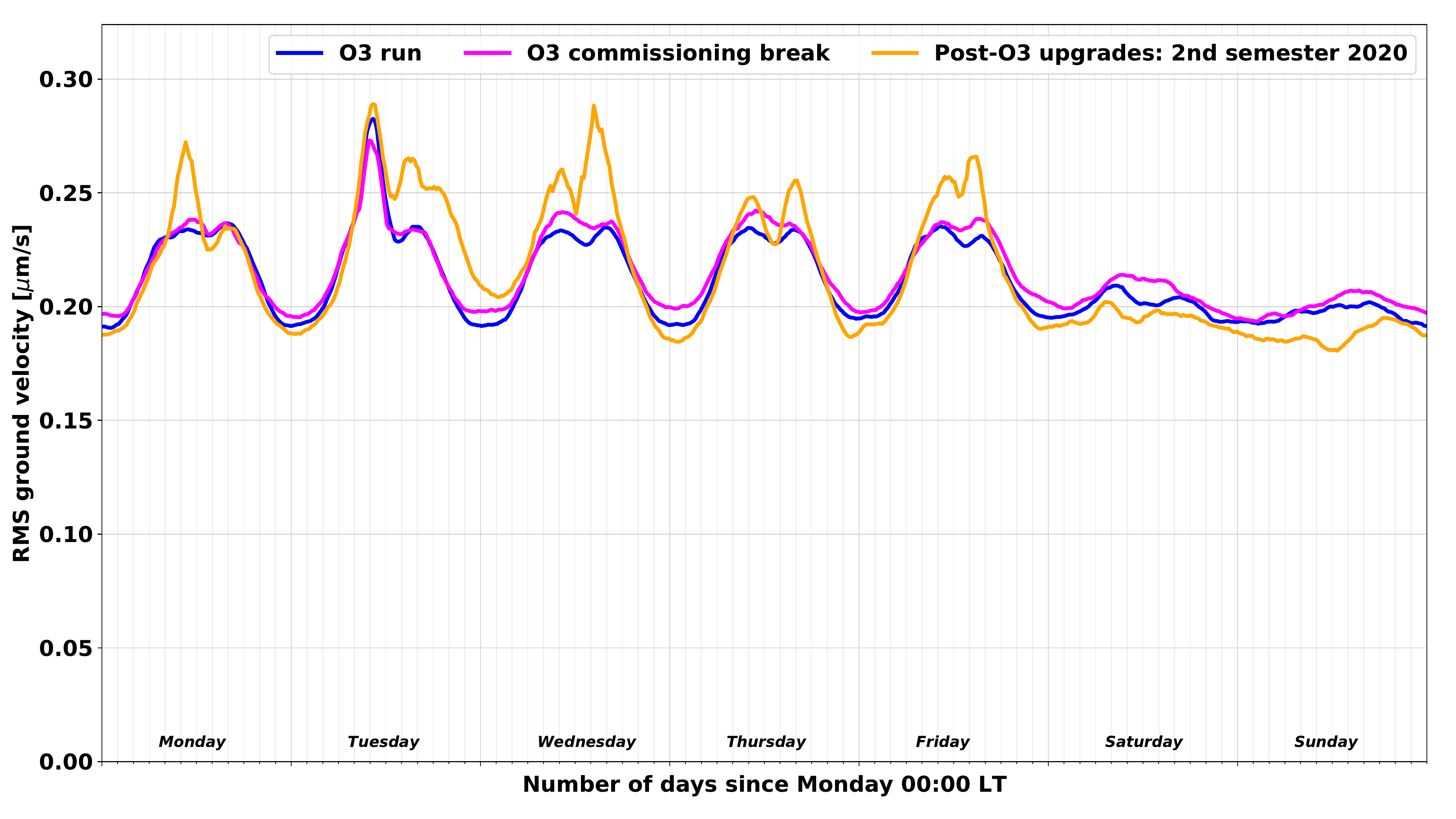}
    \end{center}
    \caption{Average evolution on a weekly basis of the seismic on-site noise (frequency band: 10-40~Hz) measured at EGO during different times in 2019-2020.}
    \label{fig:onsite}
\end{figure}

Finally, above 10~Hz, the dominant seismic contribution is generated locally: vehicles in nearby and on-site roads, agricultural work on neighbouring land, etc. Figure~\ref{fig:onsite} shows the average day-night variations, computed in the 10-40~Hz band on a weekly basis: in blue during the O3 run; in magenta during the 1-month commissioning break (October 2019) separating the two halves of O3; finally in orange for the second semester of 2020, during which hardware upgrades and construction or infrastructure works for the Advanced Virgo+ project~\cite{AdV+} took place.

The common feature between the three curves is the dominant peak on Tuesday mornings, the usual slot used for the weekly maintenance of the Virgo detector. This activity includes in particular the refilling of Nitrogen\footnote{Liquid Nitrogen is used to cool down the Advanced Virgo cryotraps~\cite{TheVirgo:2014hva}.} tanks by heavy trucks coming on-site, and the possibility to have people moving around and working inside experimental areas whose access is forbidden during data taking periods. The on-site seismic noise level was slightly higher during the commissioning break compared to the O3 run, but not by much: that 1-month shutdown was not long enough to allow for invasive works that could have jeopardized the restart of data taking on November 01, 2019, alongside the two LIGO detectors. On the other hand, on-site activites are more evenly distributed over working days during the post-O3 upgrade. Though, activities were the lowest on weekends during that period because of site access restrictions, enforced because of the pandemic.

\subsection{Impact on the Virgo detector}

The previous sections have demonstrated that the Virgo collaboration is accurately monitoring the seismic environment at EGO and that the recorded data show significant variations over time, in agreement with expectations from known noise sources. It is then interesting to see how these noises impact the performance of the Virgo detector, namely its sensitivity and duty cycle.

\subsubsection{Sensitivity}
\label{section:impact_sensitivity}

A convenient way to monitor the sensitivity of a GW detector like Virgo is to study the evolution of the {\em BNS range}, that is the average distance up to which the merger of a standard BNS system can be detected with a signal-to-noise ratio (SNR) set to 8, roughly corresponding to one false alarm per year with purely Gaussian noise. The average is taken over the position of the BNS in the sky and over the orientation of its orbital plane. Broadly speaking, the lower (higher) the noise in the frequency band of interest~--from a few tens of Hz to a few hundreds of Hz depending on the actual sensitivity curve~--, the larger (smaller) the BNS range.

\begin{figure}[!htbp]
    \begin{center}
      \includegraphics[width=0.98\columnwidth]{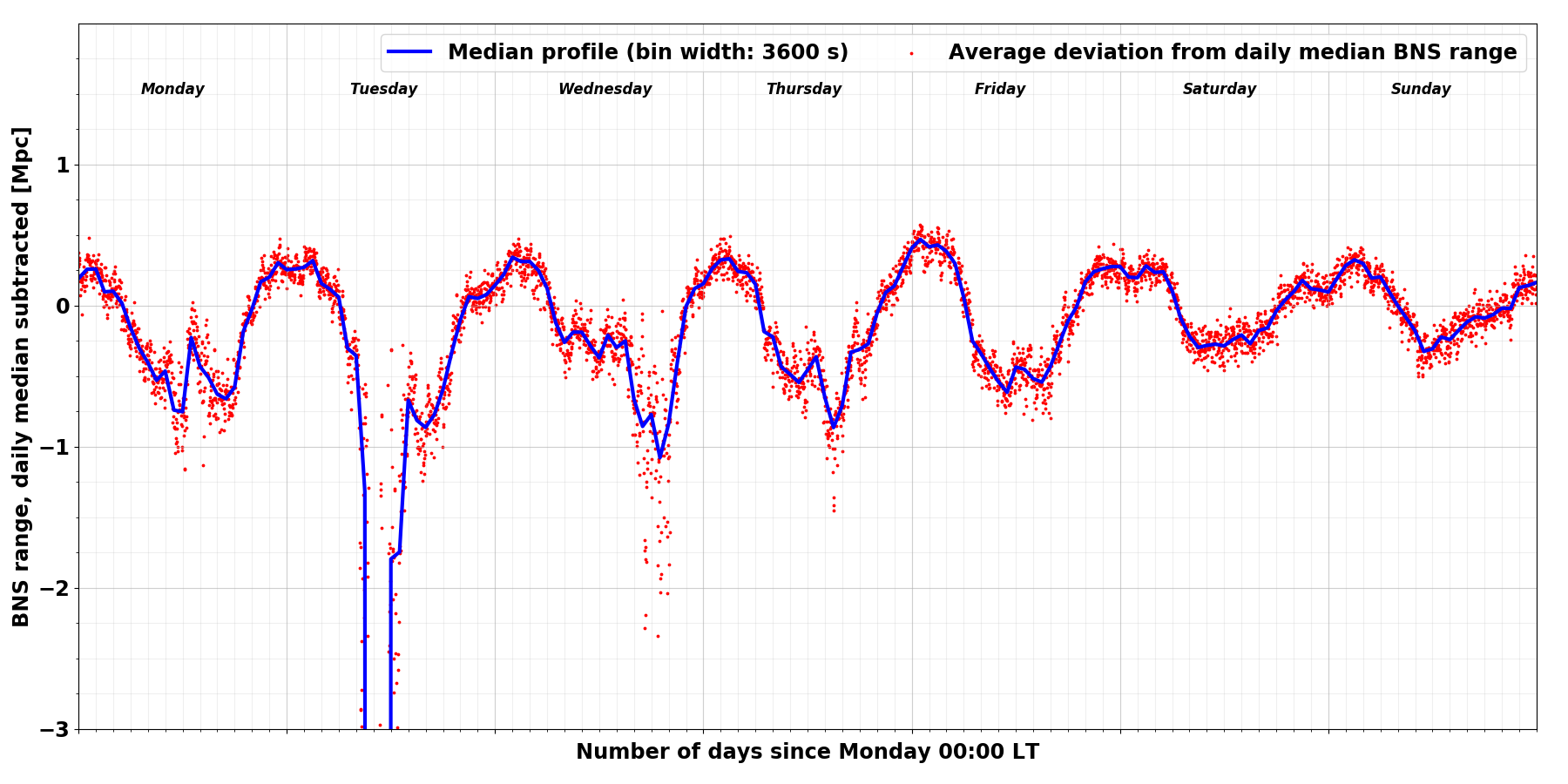}
    \end{center}
    \caption{Average variation of the BNS range around its local average, computed on a weekly basis. The blue trace is a moving median profile of the red scatter plot, each dot showing the fluctuation at a particular weekday and time. The lack of available data on Tuesday morning corresponds to the weekly maintenance period of the Virgo detector, while the sharper variations on Wednesday and Thursday afternoons are due to the fact that these times have often been used for calibration or detector activities. Therefore, the BNS range is less stable than usually when nominal data taking gets restored.}
    \label{fig:BNS_range_weekly_variations}
\end{figure}

\begin{figure}[!htbp]
    \begin{center}
      \includegraphics[width=0.98\columnwidth]{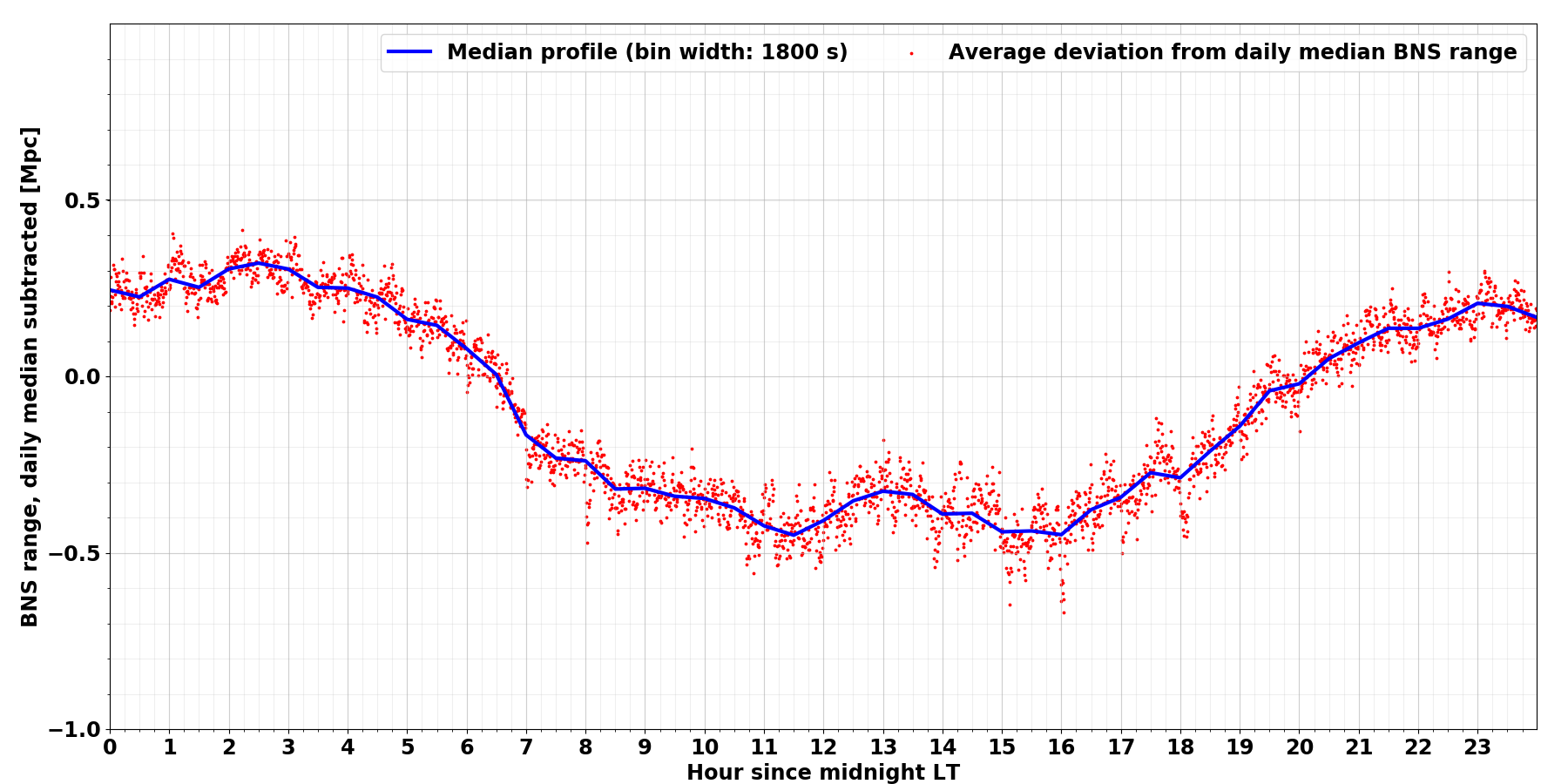}
    \end{center}
    \caption{Average variation of the BNS range around its local average, computed on a daily basis. The blue trace is a moving median profile of the red scatter plot, each dot showing the fluctuation at a particular time of the day.}
    \label{fig:BNS_range_daily_variations}
\end{figure}

In addition to its potential dependence on the surrounding environment, the BNS range can fluctuate significantly due to changes in the control accuracy of the detector. Therefore, averaging raw BNS range values, especially over long timescales, is not expected to provide meaningful information as one would mix together too many effects that cause the BNS range to vary. Therefore, the method used in the following consists in computing a moving daily average of the BNS range and to focus on the local fluctuations around this level. Figures~\ref{fig:BNS_range_weekly_variations} and~\ref{fig:BNS_range_daily_variations} show these variations, averaged over the whole O3 run, and projected over a weekly or daily time range, respectively. On both plots, the red dots show daily variations while the blue curve is a moving median profile of the scatter plot. The variations seen are clearly of anthropogenic origin, with a day-night pattern and a reduced spread during the weekend. Although they are significant, they are also limited in size: ${\sim} 1$~Mpc compared to an average BNS range of about 50~Mpc during the O3 run, hence a ${\sim}2\%$ fluctuation. This shows the robustness of the Virgo detector.

\subsubsection{Duty cycle}

\begin{figure}[!htbp]
    \begin{center}
      \includegraphics[width=0.98\columnwidth]{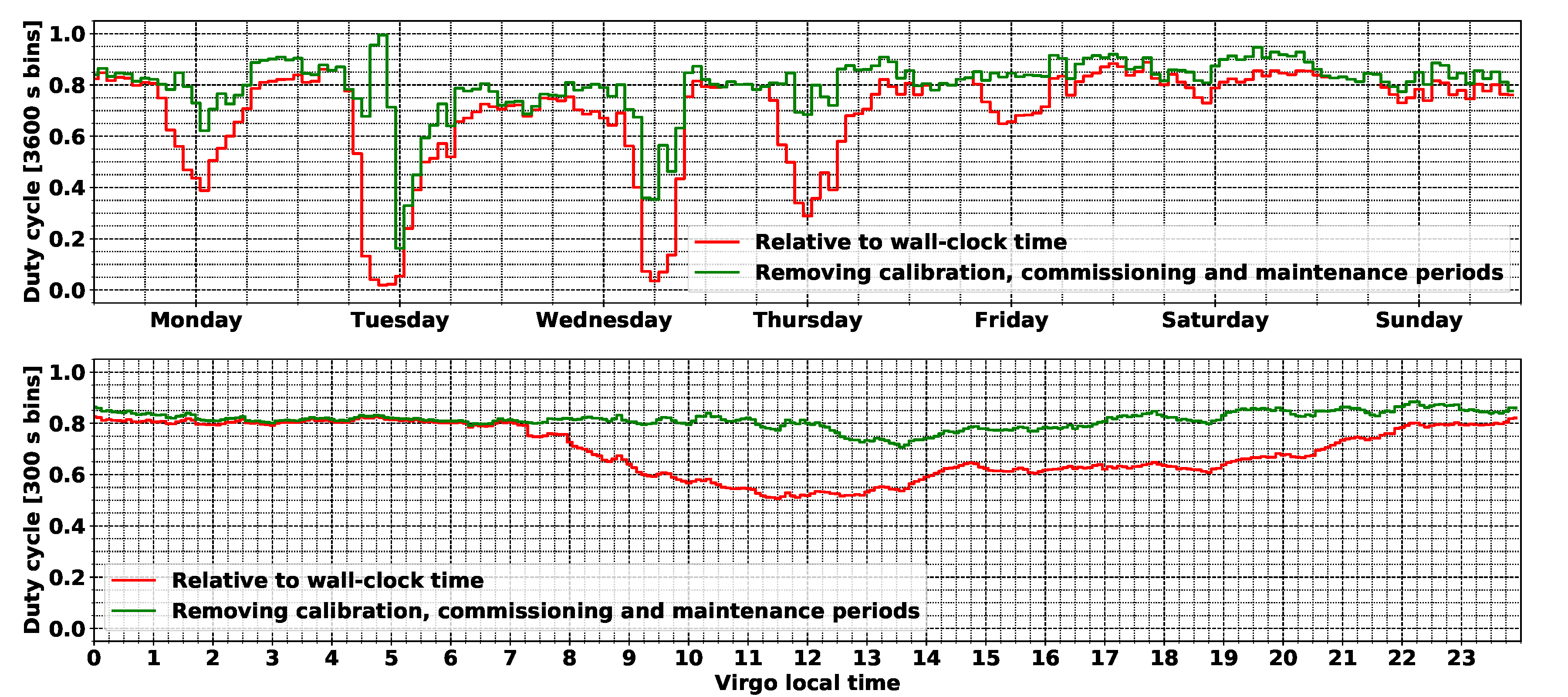}
    \end{center}
    \caption{Average weekly (top) and daily (bottom) duty cycle of the Virgo detector during the O3 run. The red curve uses the elapsed real time as normalization, while the green one is produced excluding the times spent doing calibration, commissioning or maintenance, three activities that are incompatible with Science-mode data taking.}
    \label{fig:duty_cycle}
\end{figure}

Figure~\ref{fig:duty_cycle} shows the average duty cycle of the Virgo detector during the O3 run. The top plot displays its average variation over a week, while the bottom one focuses on a day. The red curve normalizes the Science mode data taking by the elapsed real time, while the green one is computed by excluding the calibration, commissioning and maintenance periods. Thus, the latter curve shows the fraction of the time available for data taking that is actually used for that. Activities on the detector are concentrated during working hours as expected, with maintenance on Tuesday morning, calibrations on Wednesday evenings and commissioning slots from Monday to Friday depending on the needs. There is a non-negligible recovery time from maintenance, while the transition from calibration back to data taking is smoother and quicker on average. During the quietest hours of the night, when no work takes place on the interferometer except in case of an emergency, the average duty cycle reaches a plateau around 85\%.

\section{Earthquakes}
\label{section:earthquakes}
\markboth{\thesection. \Sectionname}{}
Earthquakes radiate energy through different types of seismic waves that are commonly
divided in "body" and "surface" waves, depending on the path followed from the source to the
receiver. Body waves that travel through the Earth are usually detected first. The fastest are
named P-waves and are compressional longitudinal waves whose speed can reach 8~km/s.
Then come the S-waves, transverse shear waves whose velocity scales by a factor of $\sqrt{2}$
with respect to P-waves. Surface waves are slower and their size dominates at large
epicentral distance since their amplitude scaling factor is 1/distance while body waves scale with 1/distance$^2$. Most relevant surface waves are Rayleigh waves, that originate from P-wave and S-wave (with vertical polarization) coupling at the Earth surface. The result is a wave with both longitudinal and transversal components and a propagation speed up to a few~km/s.

Since seismic waves excite buildings even at great distance, the Virgo monitoring set at EGO includes local, regional and teleseismic earthquakes since it was observed that all of them can induce large motion
of the interferometer elements. This can then saturate the control capability of the feedback systems that keep Virgo at its nominal working point, leading to a loss of control. Following each control loss (regardless of its origin: an earthquake or another cause), data taking stops immediately and can only restart after the completion of the semi-automated sequence that allows restoring the Virgo global working point~-- during the O3 run, that procedure took about 20 minutes on average~\cite{o3virgodetchar}. But the time lost can be much longer in case of a control loss due to an earthquake, in case the suspension normal modes are excited by the seismic waves. In that case, one may have to wait up to one hour after the event that the high-quality factor modes of the suspensions are damped, before initiating the control acquisition procedure. Since each control loss reduces the Virgo duty cycle, it is therefore important to understand which fraction of these are due to earthquakes, and what are the earthquakes that induce them.

Large earthquakes at local and regional scale do not happen very often, so the type
of earthquakes on which this analysis is focused is large
earthquakes that occur along the boundaries of the main tectonic plates. Most of them are quite distant from EGO, meaning that a low-latency framework relying on data from a variety of seismic stations worldwide could produce early warning notices that would be received and processed {\it ahead} of the seismic waves arrival. In that case, one could take preventive measures to try to mitigate the effect of the ground shaking, with the goal of avoiding the control loss. In the following, we describe the strategy implemented at Virgo during the O3 run, that relies on the Seismon framework~\cite{Coughlin_2017,Biscans_2018,Mukund_2019} developed at LIGO~-- an example of the existing teamwork among members of the global GW detector network. 

Furthermore, as explained below, the study has also highlighted another contribution from much weaker earthquakes, quite close to EGO (the majority of which occur on the Italian Apennines). Those have been more difficult to identify as they do not lead to early warnings from Seismon and the frequency of their seismic waves is much higher when they arrive at EGO: up to ${\sim} 1$~Hz, whereas teleseism waves are in the frequency band $10-100$~mHz. In addition, the proximity of their epicenters makes useless the use of warnings that would always come too late. Thus, the only way to mitigate these earthquakes is to understand how they impact the Virgo control system and what could be done to strengthen it.

\subsection{O3 Seismon setup at EGO}

In addition to making the whole detector as robust as possible against the passing of strong seismic waves, the only other leverage one can use to mitigate the impact of earthquakes is to rely on early warnings provided by worlwide arrays of seismometers.

Following initial tests done during the O2 run and the upgrade period that followed, we ran at EGO during O3 an instance of the Seismon framework, developed by LIGO to process earthquake early warnings provided by the US Geological Survey (USGS)~\cite{USGS_PDL} and to compute information relevant for each site of the LIGO-Virgo network.
Namely, for each earthquake, Seismon potentially predicts the arrival time of the different
seismic waves (P-, S- and surface), their amplitude at site and the probability of losing the detector
control in consequence of that earthquake
That framework was split into four consecutive steps, each associated with a server integrated within the Virgo online data acquisition system (DAQ) used to steer and monitor the detector.

\begin{itemize}
\item Reception of the USGS alerts.
\item Processing of each alert by the Seismon framework.
\item Extraction of the subset of Seismon data pertinent to the EGO site and provision to the Virgo online framework.
\item Local processing of these data.
\end{itemize}

In addition to producing a plot summarizing all information available from the early warning, a loose cut is applied on magnitude and distance to estimate whether the earthquake could be relevant, meaning that it could impact the control of the Virgo detector. In that case, and if the warning was quick enough to precede the arrival of the seismic wave on-site, an alarm would latch on the main panel of the Virgo Detector Monitoring System~\cite{dms1,dms2}, alerting the operator on duty in the control room.

In the nominal O3 control configuration, the two 3~km-long optical cavities are kept in resonance by acting on the end mirror suspensions: their actuators are the least noisy, at the price of a reduced correction range availability. Actuators located at the level of the input mirror suspensions have higher dynamics, while introducing slightly more noise. Thus, they can be used as earthquake control mode (in short {\em EQ mode}) to try to maintain the Virgo working point during periods of elevated seismic noise. 

A smooth transition procedure, working both ways without losing the detector control, has been developed to allow switching back and forth between end-mirror and input-mirror actuations. During the O3 run, the procedure in use was the following: once alerted by Seismon, the operator on duty would monitor the optics suspension status and manually trigger the transition from nominal mode to EQ mode when the test mass suspensions would start shaking significantly. Once activated, that process would take a few tens of seconds to complete. Then, either the detector would nevertheless lose its working point (and the control acquisition procedure would have to be restarted from the beginning), or the EQ mode control would be kept until the whole seismic wave trains has passed by and the suspensions motion has been damped back to levels low enough to allow resuming the nominal control mode.

Unrelated to earthquakes, the EQ mode was also found useful during periods of high wind: gusts shake the building structures (walls and floors) and those vibrations can couple to the suspensions, potentially causing control corrections to saturate.
However, since EQ mode was not validated for the production of good quality data for physics analysis, this method was used parsimoniously during most of O3 because corresponding data would have to be discarded.
A few weeks before the end of the run, the EQ mode got finally qualified for regular data taking and later studies~\cite{logbook_rolland_EQ_mode_OK} showed that there was no significant degradation of the Virgo sensitivity when switching to it. Therefore, it was used more regularly from that time; the possibility to have such a backup solution for O4 as well will be studied in the coming months.

\subsection{Earthquakes impact during O3}

The stronger and/or the closer to EGO the earthquake, the more likely it is to impact the control of the Virgo detector. To study the impact of strong regional earthquakes or teleseisms, the USGS warnings processed by Seismon are sufficient (as they should include all such earthquakes). But it was soon realized that some 
moderate earthquakes occurring at local and regional distance
(from few tens to few hundreds kilometers away from EGO), too weak to generate a USGS alert and thus not processed by Seismon, could cause losses of control of Virgo.
To check if any of the control losses
was caused by this type of earthquakes, we queried~\cite{INGV_website_query} the INGV (Istituto Nazionale di Geofisica e Vulcanologia) public earthquake catalogue~\cite{INGV_website} to
download the list of events that occurred during O3 in the Mediterranean region. This list partly
overlaps with the USGS one and duplicates were removed. All results presented in the following are based on the whole set of earthquakes, assembled by merging the USGS and INGV event lists. 

The control of the Virgo detector is extremely complex.
Therefore, finding out how many earthquakes induced control losses during the O3 run required a careful study of all control losses, documented below in \ref{section:lock_losses}. An earthquake from the list of USGS warnings is associated to a recorded control loss if the loss occurs within the time range during which seismic waves were predicted to arrive on-site according to Seismon and if the seismic activity around the time of the control loss is significantly larger than its typical range of variation. In case of concurring early warnings from different earthquakes overlapping in time at EGO, the strongest is arbitrarily selected as the reason for the control loss.

Estimating the strength of an earthquake when its seismic waves arrive at EGO is not easy. Yet, this is a key point to address, first to reject quickly warnings from harmless earthquakes and then to adjust the latency and level of response for the crew in charge of steering the Virgo detector. During O3, basic rectangular cuts in the magnitude-distance plane~-- e.g. {\it if magnitude > (...) or (distance < (...) km and magnitude > (...)) or etc.}~-- were applied to the live earthquake warnings received from USGS and processed by Seismon. During the post-run analysis, the ranking

\begin{equation}
\mathrm{ranking} = \frac{10^{\mathrm{magnitude}/2}}{\mathrm{distance [km]}}
\label{eq:EQ_ranking}
\end{equation} 

was introduced. While not complete~-- e.g. neither the hypocenter depth nor its azimuth angle computed with respect to EGO are accounted for~-- this ranking appears sound: the higher its value, the more likely the control loss.
Applying a (conservative) minimum cut at ranking~=~0.02 allows to safely
remove more than half of the earthquakes to be analyzed.

Results shown below use the largest possible earthquake statistics,
meaning that one requires the Virgo detector to be fully controlled,
but not necessarily in Science mode. This looser requirement enlarges the dataset of interest and hence the number of earthquake early warnings to be taken into account.

\begin{figure}[!htbp]
    \begin{center}
      \includegraphics[width=0.98\columnwidth]{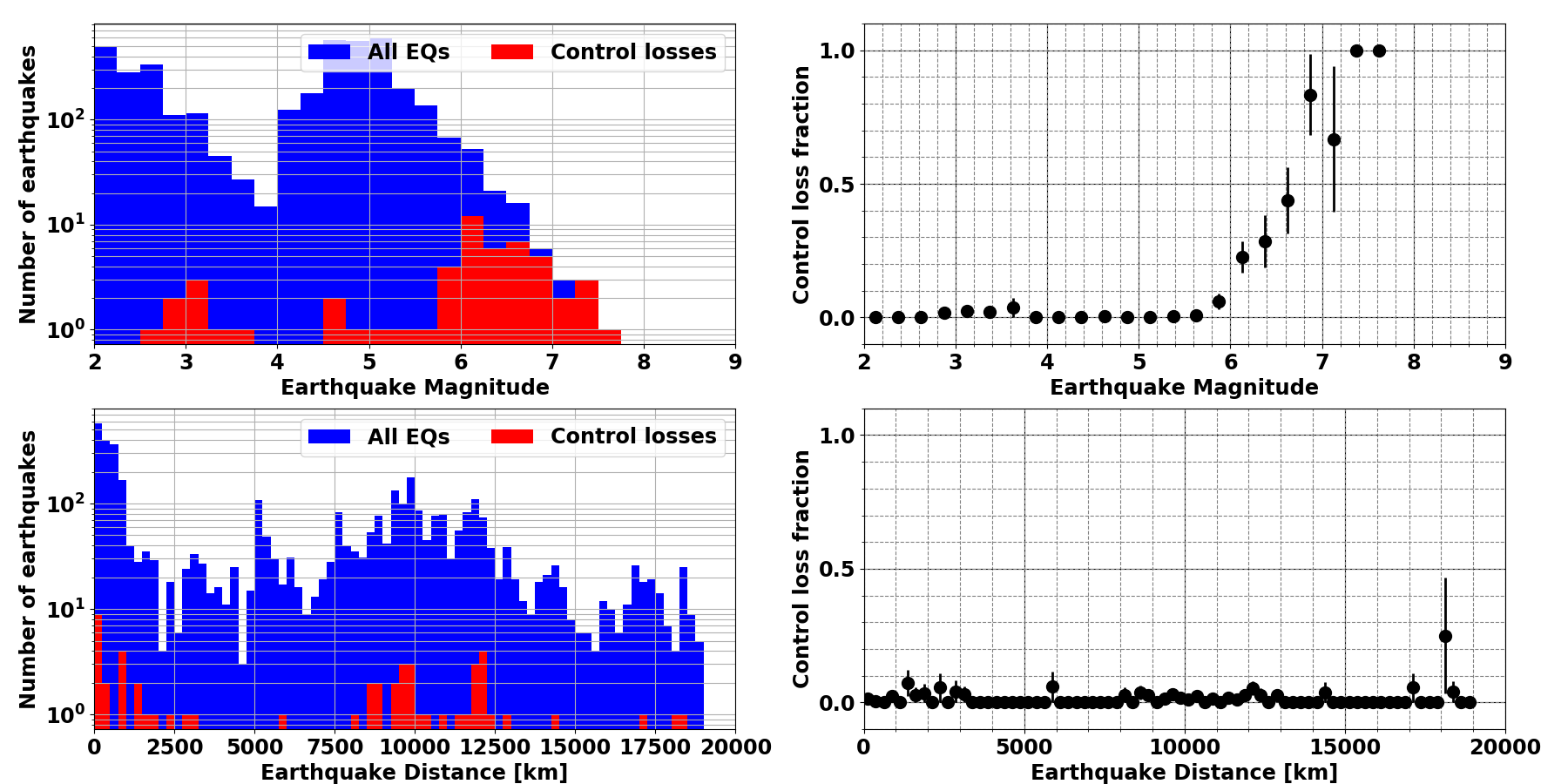}
    \end{center}
    \caption{Summary of the impact of earthquakes on the Virgo detector during the O3 run. Left column: the blue (red) histogram shows all earthquakes (the earthquakes that have induced a control loss); top: magnitude distribution; bottom: distribution of the distance between EGO and the epicenter. Right column: corresponding fraction as a function of the earthquake magnitude (top) and distance (bottom). In all cases, the earthquakes that certainly could not impact Virgo (ranking below 0.02) were excluded.}
    \label{fig:EQs_magnitude_distance}
\end{figure}

Figure~\ref{fig:EQs_magnitude_distance} highlights the epicentral distance and magnitude of the earthquakes that led to a Virgo control loss. The top (bottom) row deals with the earthquake magnitude (epicentral distance) while the right column displays the ratio of the red and blue histograms shown on the left column.
As expected, the larger the earthquake magnitude, the more likely the control loss, with the fraction of earthquakes leading to a control loss departing from 0 for magnitude 6 and above. That fraction saturates to 1 (meaning that all events causes a control loss) when magnitude exceeds 7.2. We also note that the fraction is not null around magnitude 3: this reflects the control loss consequence of some small local earthquakes recognizable also in the left side
histogram of Figure~\ref{fig:EQs_magnitude_distance}.
The histogram ratio is much flatter for that other variable, with the most significant bins reflecting the location of seismic regions on the globe with respect to EGO, mainly the broad Mediterranean area and the Ring of Fire (a region covering much of the rim of the Pacific Ocean that is seismically very active).

\begin{figure}[!htbp]
    \begin{center}
      \includegraphics[width=0.98\columnwidth]{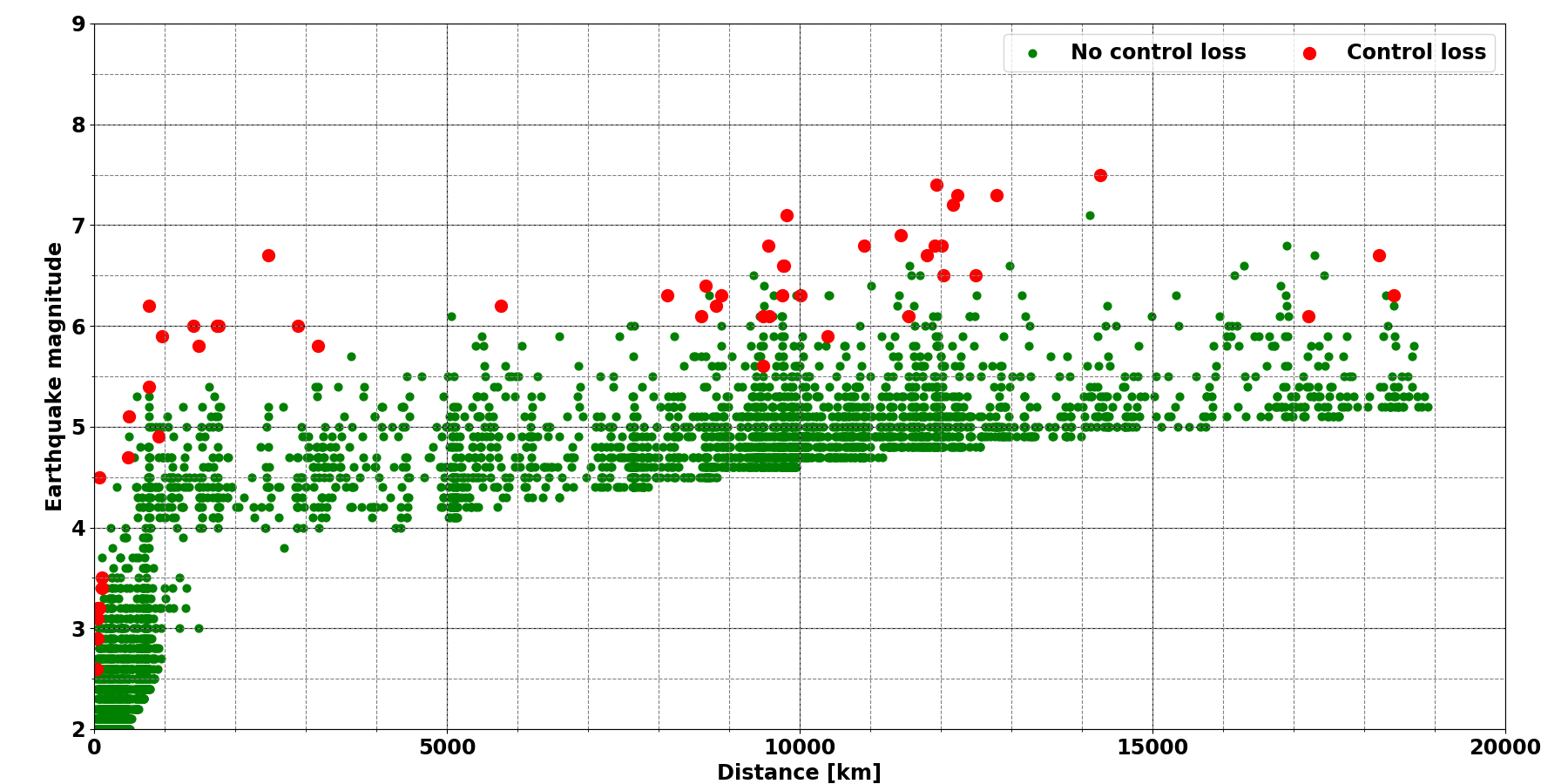}
    \end{center}
    \caption{Distribution of earthquakes in the plane distance-magnitude during the O3 run. The earthquakes that caused a control loss (did not cause a control loss) are represented with red (green) dots. The lack of points below the main bulk of earthquakes is due to the ranking cut, set at 0.02.}
    \label{fig:EQs_magnitude_distance_2D}
\end{figure}

Figure~\ref{fig:EQs_magnitude_distance_2D} shows the population of earthquakes that caused a control loss (red dots) in the two-dimensional plane epicentral distance vs. magnitude. These earthquakes form the upper envelope of the scatter plot drawn, meaning they are usually the 
earthquakes with largest magnitude for any distance.
The separation between red and green (earthquakes that did not cause a control loss) dots is not perfect for at least two reasons. The first one is that the control of the Virgo detector is complex enough that the actual level of control (accuracy and stability) plays a role in whether or not the control is lost for earthquakes at the limits of inducing a control loss. The second reason is that our model could probably be improved by including other earthquake warning parameters: two candidates would be the hypocenter depth (the deeper the hypocenter, the lower the earthquake impact on the ground at equivalent magnitude) and the azimuthal orientation of the epicenter with respect to EGO.

\begin{figure}[!htbp]
    \begin{center}
      \includegraphics[width=0.98\columnwidth]{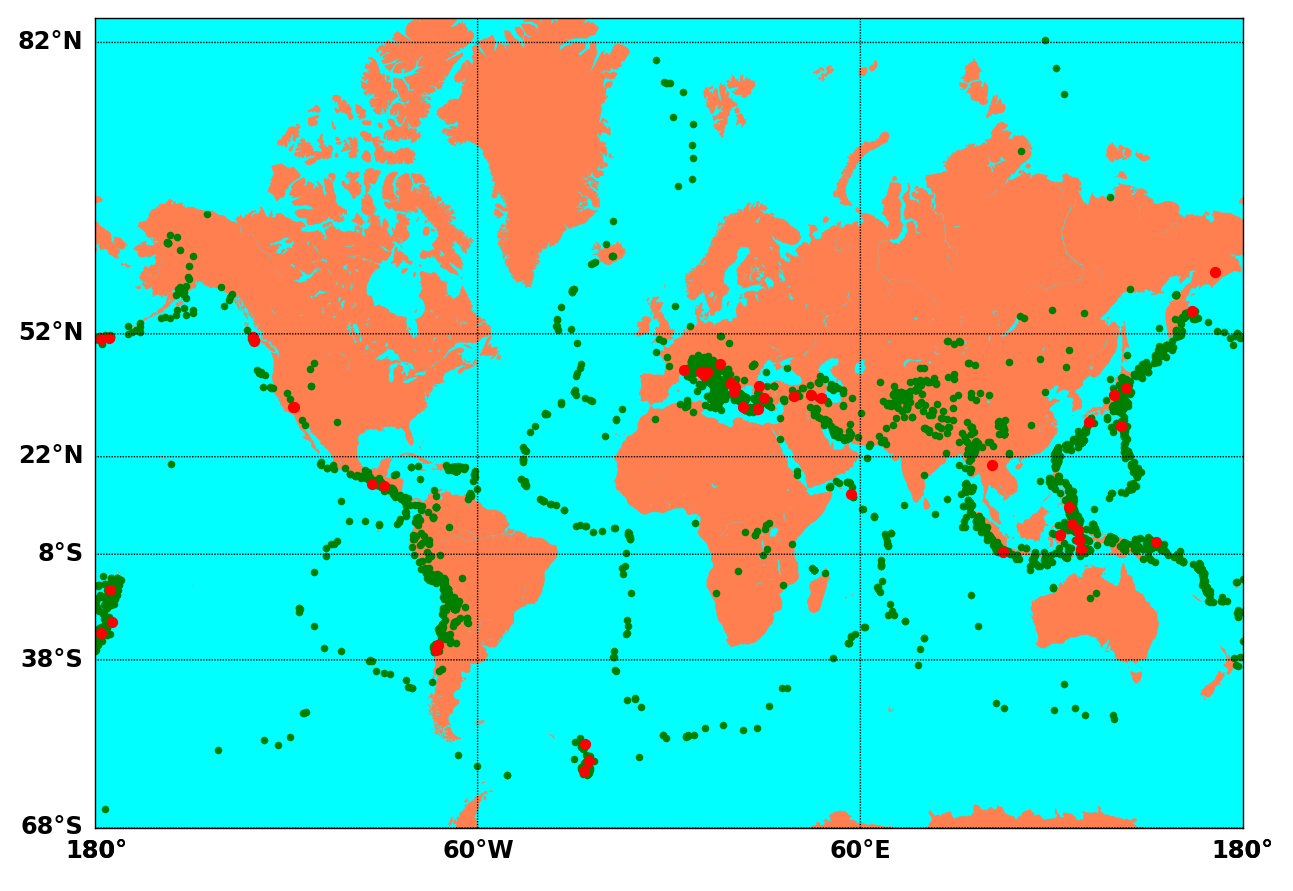}
    \end{center}
    \caption{
Location of the O3 earthquakes used in this study (ranking greater than 0.02)
The earthquakes that caused a Virgo control loss (did not cause a control loss) are represented with red (green) dots.}
    \label{fig:EQs_Earth}
\end{figure}

\begin{figure}[!htbp]
    \begin{center}
      \includegraphics[width=0.98\columnwidth]{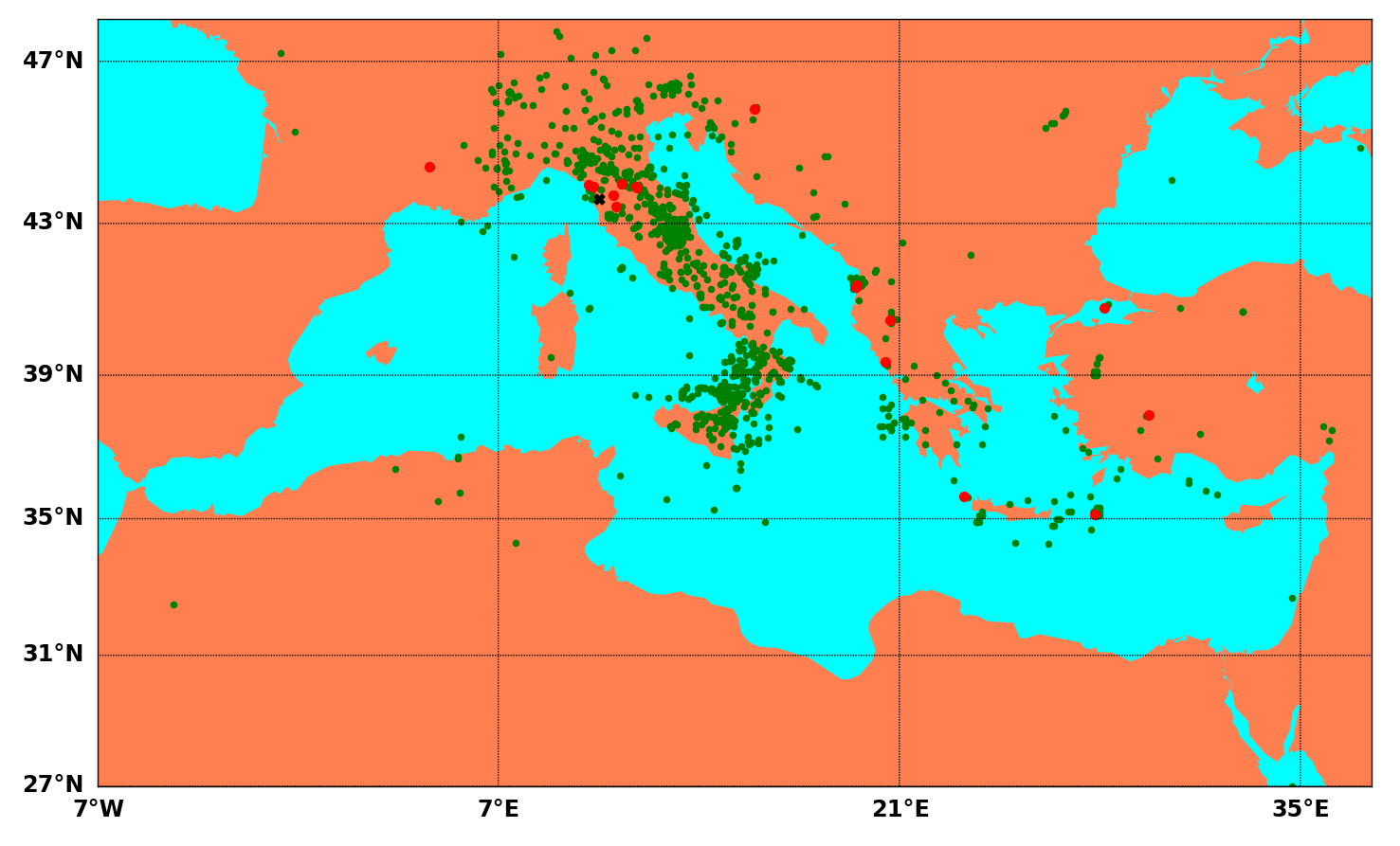}
    \end{center}
    \caption{Zoom on the Mediterranean area of the map shown in Fig.~\ref{fig:EQs_Earth} above. It shows the earthquakes nearby Virgo (whose site, EGO, is marked by a black cross) recorded during the O3 run. The earthquakes that caused a control loss (did not cause a control loss) are represented with red (green) dots.}
    \label{fig:EQs_Mediterranean_area}
\end{figure}

Figures~\ref{fig:EQs_Earth} and~\ref{fig:EQs_Mediterranean_area} show the 
location of the significant earthquakes that occurred during 03 with the same color coding used in Figure~\ref{fig:EQs_magnitude_distance_2D}. Their distribution depicts the boundaries of the main tectonic plates and, as discussed above, we can observe that 
the most harmful earthquakes for Virgo are coming from the Mediterranean area (medium to large magnitudes but smaller distances) and part of the Pacific Ring of Fire. The mid-Atlantic ridge and the Asian portion of the Alpide earthquake belt did not produce many earthquakes that impacted Virgo, possibly because of the limited statistics. During the O3 run, the distribution of the earthquakes leading to control loses was the following: ${\sim} 15$\% of close earthquakes; ${\sim} 20$\% from other earthquakes in the Mediterranean area; and ${\sim} 65$\% from distant earthquakes.
We remark that this statistics has not an absolute meaning: the O3 run took place during a quiet seismic period for Italy, compared to e.g. 2009 or 2016.
This analysis will be updated in the future with data from the O4 run.

\begin{figure}[!htbp]
    \begin{center}
      \includegraphics[width=\columnwidth]{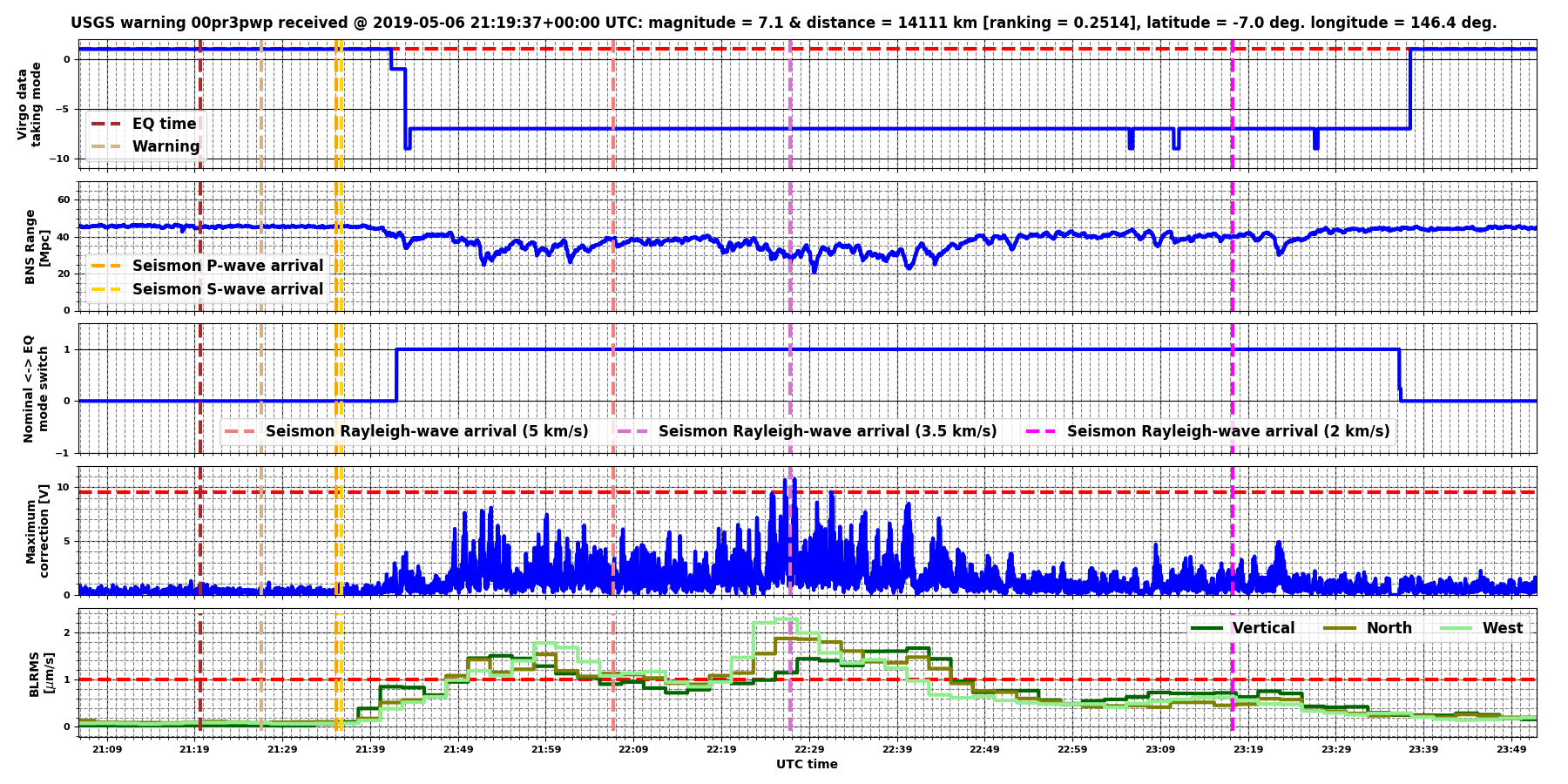}
    \end{center}
    \caption{Example impact on the Virgo detector of a strong (magnitude greater than 7) and distant (14,000 km away from EGO) earthquake, that occurred on May 06, 2019 at 21:19:37 UTC in Eastern Papua New Guinea. The description of the different stripcharts is provided in the text.}
    \label{fig:00pr3pwp}
\end{figure}

Finally, Fig.~\ref{fig:00pr3pwp} shows an example of the impact of a strong and distant earthquake on the Virgo detector and how the early warning information was used to change the control mode prior to the arrival of the strongest seismic waves. This allowed the crew on duty to keep the working point of the instrument by preventing the correction force (applied on mirror suspensions to maintain resonance in the arm cavities) from saturating. Should that action not have been performed, the control loss would have been unavoidable~-- as the correction would have saturated around 22:27 UTC. The description of the different stripcharts displayed is given below.

\begin{itemize}
\item Top plot: variation of the index labelling the Virgo data taking configuration: the Science mode corresponds to the value 1; other indices shown here (-1, -7, -9) indicate different control configurations that are not nominal and that were used to wait for the right moment to switch back to Science data taking mode.
\item Second plot: stripchart of the BNS range versus time; the seismic waves clearly make the BNS range go down and fluctuate more while they are passing (see seismic activity variations recorded in the bottom plot, described below); the BNS range recovers its steady value at the end of the plotted time when the earthquake effect fades away.
\item Third plot: switch showing the times when the earthquake-resilient control mode ('EQ-mode') is turned on ($0 \rightarrow 1$ transition) and later on off ($1 \rightarrow 0$ transition) manually by the operator on-duty.
\item Fourth plot: For each second, maximum value of the correction applied on the test masses to keep the Virgo arms in resonance. When the nominal control mode is used, a control loss happens within two seconds at most after the time for which the correction voltage\footnote{The mirror control is done by varying the amount of current applied to actuators (pairs of coil-magnet): see Ref.~\cite{ACERNESE2020102386} for details.} exceeds a 9.5~V threshold. This occurs a few times close to the middle of the time range represented here but no control loss follows, as the EQ-mode allows for larger corrections.
\item Bottom plot: seismic noise measured in three orthogonal directions (vertical and along the two Virgo arms) using the 
dominant frequency range for earthquakes recorded at teleseismic distance:
10~mHz $\rightarrow$ 100~mHz.
\item Finally, the vertical dashed lines common to all plots show the time of important events. From left to right: the time at which the earthquake occurred; the time at which the corresponding USGS warning had been received and processed by the Seismon framework at EGO; the expected arrival time of the seismic
P-waves, S-waves and Rayleigh waves. For the latter, we use three different arrival times that stem from different assumed velocities
(5, 3.5 and 2~km/s respectively).
\end{itemize}

\subsection{Plans for O4}

Work is in progress to build on the O3 experience and have a more performing, better integrated, earthquake early warning framework for O4 (and beyond). The plan is to run the latest version of Seismon with an improved prediction capability for EGO, achieved by means of all the data collected during the O3 run. We are also exploring the possibility to use the INGV Early-Est system (a framework for rapid location and seismic/tsunamigenic characterization of earthquakes)~\cite{nhess-15-2019-2015,EarlyEst_website} as an additional source of warnings, complementary to USGS. Tests are in progress to have this new live stream received at EGO and integrated into the existing framework. The two sets of early warnings will then be compared, in terms of latency and accuracy.

\section{Bad weather}
\label{section:bad_weather}
\markboth{\thesection. \Sectionname}{}
Through O3, the Virgo interferometer performed worst during days with adverse meteorological conditions, namely high winds and intense sea activity. 
These periods were generally associated with increased non-stationary noise in the GW signal below about 100~Hz and with some difficulties in maintaining the interferometer in its controlled state, resulting in reduced duty cycle. 
In the following, we study the impact of the increased microseimic noise associated to sea waves, then the influences of wind on BNS range, as well as the effect of wind gusts on the global interferometer controls. Because of the wind action on the sea surface, high winds and rough sea often occur together. We use a statistical approach to disentangle their effects on the detector.

\subsection{Impact of sea activity}\label{sec:seaimpact}

Microseism amplitude
at EGO increases by more than one order of magnitude between calm and rough sea periods. For $10\%$ of the time during O3, ground RMS velocity between 0.1~Hz and 1~Hz was above $4~\mathrm{\upmu m/s}$, as shown in Fig.~\ref{fig:microseism_rms}.
This happened in particular in correspondence of the seasonal change in the first part of O3b and for some periods of adverse weather conditions in the first months of 2020.
Periods of intense sea activity were associated to larger than usual strain residual noise whose characteristics and origin require further analysis.

\subsubsection{Microseism impact on strain noise}

\begin{figure}[!htbp]
	\begin{center}
		\includegraphics[width=\textwidth]{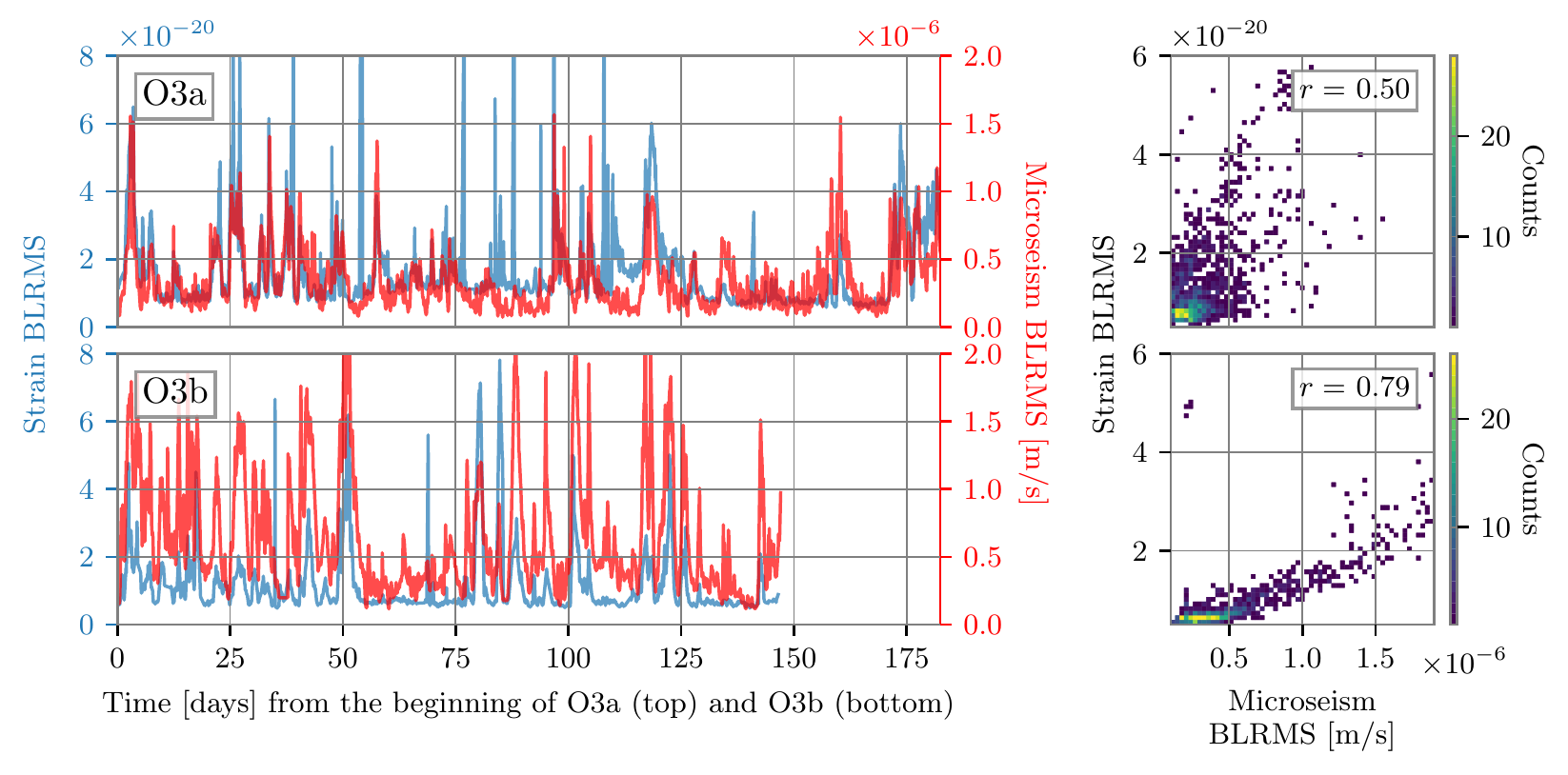}
	\end{center}
	\caption{Correlation between the low frequency noise in Virgo GW strain and the microseism induced by the sea activity; top row for O3a and bottom for O3b. {Left}: in blue the time series of the strain BLRMS in band $[10,20]~\mathrm{Hz}$ and in red that of the BLRMS in band $[0.1,1]~\mathrm{Hz}$ of a CEB seismometer, mostly influenced by the sea activity. {Right}: 2D-histograms of the correlation between the two BLRMS, where the colorscale counts, for every pixel in this map, how many data points have the corresponding values of strain and microseism BLRMS. The annotation in the top-right corner reports the value of the Pearson correlation coefficient $r$.}	\label{fig:microseism_vs_BLRMS}
\end{figure}
Periods of high sea activity were associated with larger strain residual noise up to about 100~Hz.
To characterize this effect, we made use of the \emph{band-limited} RMS (BLRMS), defined for a generic signal, in a certain frequency band $[f_\mathrm{min}, f_\mathrm{max}]$, as:
\begin{equation}\label{eq:BLRMS_def}
\mathrm{BLRMS}\big(t;[f_\mathrm{min},f_\mathrm{max}]\big):=\sqrt{\int_{f_\mathrm{min}} ^{f_\mathrm{max}} S(f;t) df}
\end{equation}
where $S(f;t)$ is an estimate of the signal \emph{power spectral density} (PSD) referred to a time $t$.

In Fig.~\ref{fig:microseism_vs_BLRMS}, we report, for the entire O3 run, in blue the BLRMS of the strain in the band $[10,20]~\mathrm{Hz}$ and, in red, the CEB seismometer BLRMS in the band $[0.1,1]~\mathrm{Hz}$.
These have been estimated from~(\ref{eq:BLRMS_def}), where $S(f;t)$ is computed with the Welch's method making use of strides of $2048~\mathrm{seconds}$ and FFT length of $128~\mathrm{seconds}$, overlapping by $50\%$~\cite{welch1967use}.
The correlation between the two curves is apparent.
In particular, when the microseism is intense, the peaks in the strain BLRMS are almost everywhere coincident with those in the seismometer BLRMS.
This fact is also highlighted in the 2D-histograms on the right-hand side of the same figure, where the Pearson correlation coefficient has been computed for the two data taking periods, O3a (top) and O3b (bottom).
In general, we observe that, despite the ``spikes'' in correspondence of bad weather conditions (in particular at the beginning of O3b and then during most of Winter\footnote{That calendar season starts around day 50 of O3b and lasts almost until the end of the data taking.}), the induced strain noise at low frequency has improved during O3 and can now be mostly attributed to microseism.

\subsubsection{Microseism impact on glitch rates}

\begin{figure}[!htbp]
	\begin{center}
		\includegraphics[width=\textwidth]{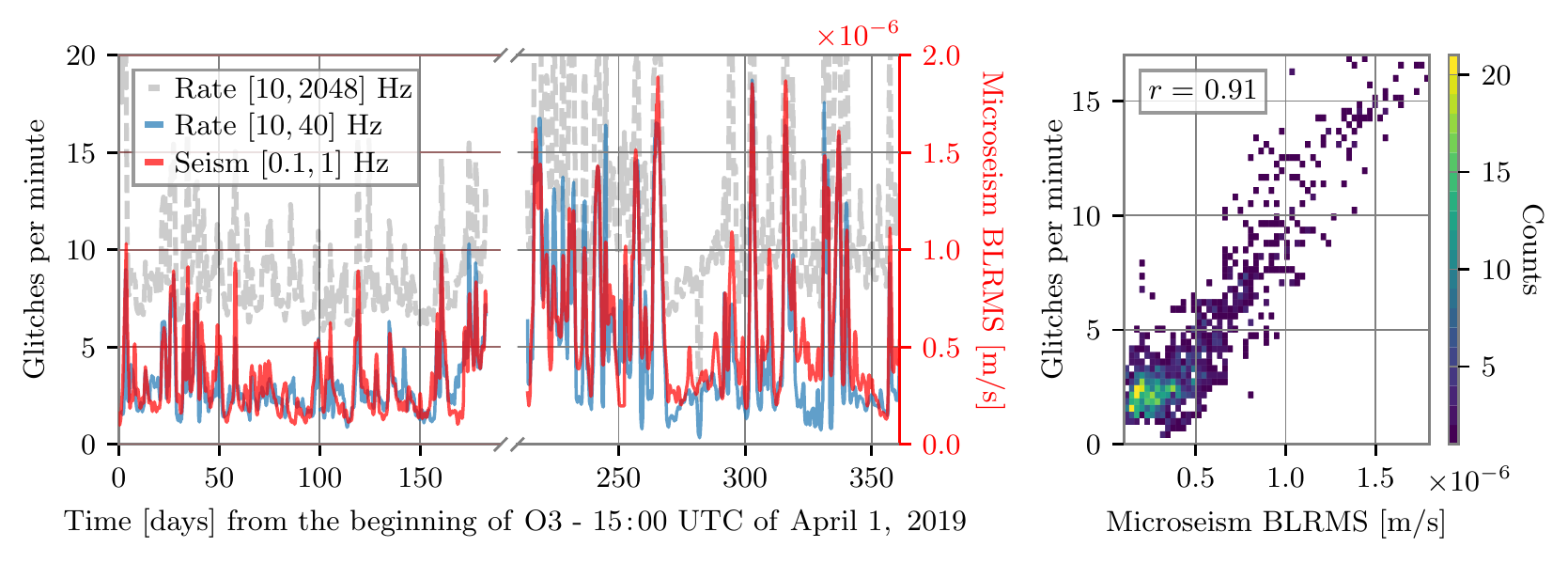}
	\end{center}
        \caption{Correlation between Virgo glitch rate and the sea induced microseism during the O3 run. Left: the dashed gray line represents the daily moving median of the glitch rate per minute recorded by Omicron~\cite{ROBINET2020100620} for glitches with SNR $>$ 6.5 and frequency at peak in band $[10,2048]~\mathrm{Hz}$, estimated over strides of $2048$ seconds. The blue continuous line is the median rate referred to glitches with frequency at peak in $[10,40]~\mathrm{Hz}$ band. The continuous red line is the BLRMS in band $[0.1,1]~\mathrm{Hz}$ of a seismometer in the Virgo CEB. Right: 2D-histogram of the glitch rate in band $[10,40]~\mathrm{Hz}$ and the microseism BLRMS, where the colorscale counts, for every pixel in this map, how many data points have the corresponding values of the rate and the microsiesm BLRMS. The annotations in the top-right corners report the values of their Pearson correlation coefficient $r$.}
	\label{fig:microseism_vs_glitch_rate}
\end{figure}

Besides an increase in the RMS value of the strain noise at low frequency, microseisms induce short transients of power excess in this channel, colloquially referred to as \emph{glitches}.
In Fig.~\ref{fig:microseism_vs_glitch_rate} we report the minute rate of these glitches during the entire O3 run.
To reduce the -- usually very large -- variability in their rate, we computed running daily medians.
The gray dashed line represents the time evolution of daily medians for glitches with SNR $>6.5$ and frequency at peak in the band $[10,2048]~\mathrm{Hz}$, as measured by the online Omicron pipeline~\cite{ROBINET2020100620}.
The blue solid line is the median minute rate of glitches with peak frequency in the $[10,40]~\mathrm{Hz}$ band. These glitches accounted for about $30\%$ of the total during O3a, and for almost $40\%$ in O3b, with peaks larger than $80\%$ in correspondence of periods of intense sea activity.
This glitch rate is highly correlated with microseism,
represented in the left-hand side plot of 
Fig.~\ref{fig:microseism_vs_glitch_rate} by the solid red line of the running weekly median of the BLRMS in band $[0.1,1]~\mathrm{Hz}$ of the CEB seismometer.
On the right-hand side of the same figure, we report the 2D-histogram of these two quantities and the value of their Pearson coefficient ($r = 0.91$).

\subsubsection{Microseism and scattered light}

\begin{figure}[!htbp]
\centering
\includegraphics[scale=0.8]{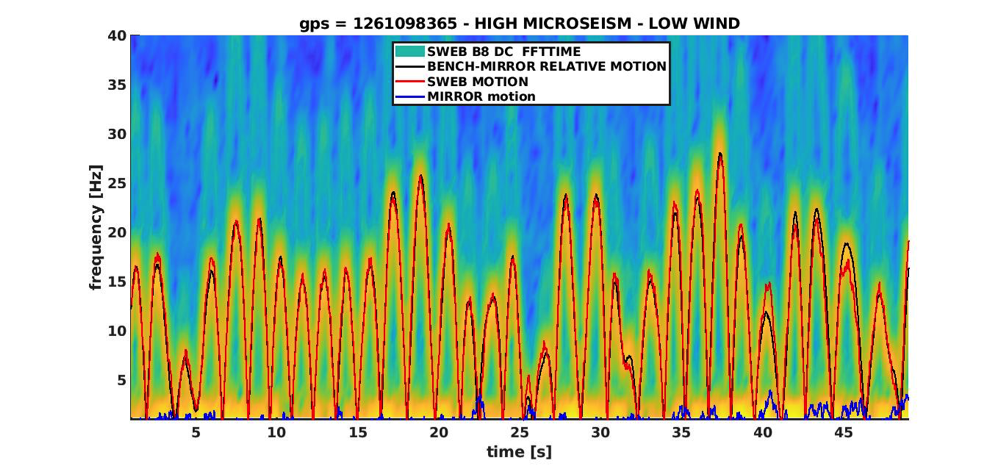}
\includegraphics[scale=0.8]{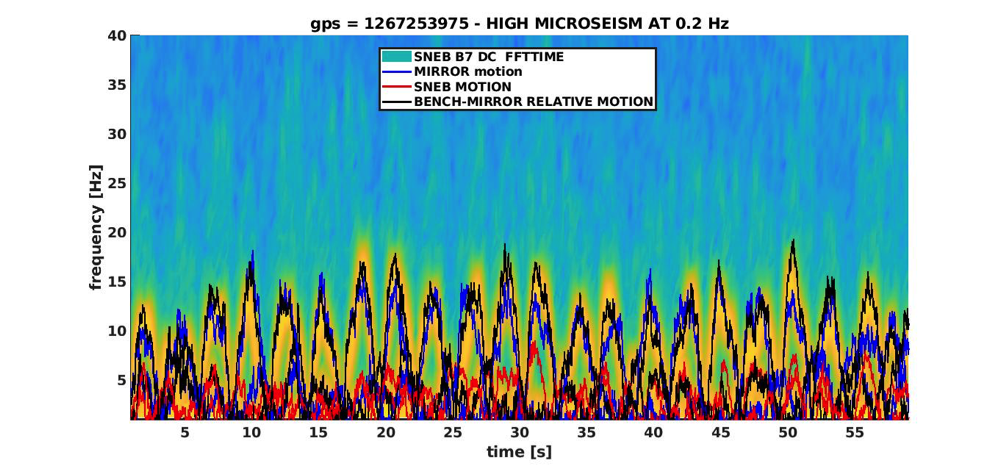}
\caption{Spectrograms of the light transmitted at the end of the arm cavities and detected by photodiodes located behind, on suspended benches~-- top plot: west arm, B8 photodiode, SWEB bench; bottom plot: north arm, B7 photodiode, SNEB bench. The typical pattern of scattered light noise (arches)~-- both first order and second order (higher frequencies)~-- is visible. On the SWEB plot, arch spacing and amplitude correspond to half the period of marine microseism at Virgo (${\sim} 3$~s) and a ground velocity of about 8~$\mu{\rm m/s}$. The predictor for BENCH-MIRROR is shown in black, while the predictors computed from mirror and bench motions are shown in blue and red, respectively. The overlap shows that BENCH-MIRROR is the best predictor of scattered light, closely matching the observed arches.}
\label{fig:spectrograms}
\end{figure}

Glitches due to microseism often resemble arches in a time-frequency map, as illustrated for example in Fig.~\ref{fig:spectrograms}. Arches are the typical signature of scattered light (SL) noise processes, which is a major issue and topic of investigation in the second generation GW detectors~\cite{Virgo_env_O3,LIGO_env_O3,Washimi_2021,Canuel:13,W_s_2021,Accadia_2010}.

A stray light beam bouncing off a moving surface adds coherently to the beam main mode every time its optical path, $x(t)$, changes (increases or decreases) by an integer wavelength. It follows that the frequency of the strain noise is:
 
\begin{equation}
f_{sc}(t)=\frac{2 n |\dot{x}(t)|}{\lambda}
\label{eq:arches}
\end{equation}

where $\dot{x}(t)$ is the instantaneous relative velocity between the interferometer beam and the scatterer, and $\lambda=1.064~\mu {\rm m}$ is the Virgo laser wavelength. Equation~\ref{eq:arches} is referred to as predictor. In case the scattered beam encounters a second reflective surface it can bounce back and forth $n$ times along the same path before recombining, giving rise to higher order noise arches, reaching out $n$-times larger frequencies. 

In O3 the main sources of SL affecting the sensitivity were the suspended optical benches placed beyond the end test masses in the terminal buildings (SNEB, SWEB). In this case, the noise observed in the time-frequency domain is well visible as power fluctuations in the cavity. The noise appears as a series of arches, where the typical non-stationarity and non-linearity of the noise is evident. Arch time spacing is the half-period of the oscillation of the mirror-bench relative motion, and arch amplitude (i.e. the maximum frequency extension of the induced strain noise) is $f_{max} = (4\pi/\lambda)AFn$ where $A$ and $F$ are the amplitude and frequency of the oscillation. If the frequency and amplitude of the oscillation are such that $f_{max} > 10$~Hz, the noise affects the GW detection frequency band.

Being those benches suspended and controlled~\cite{SBE}, their motion induced by the microseism was supposed to be attenuated enough to push the maximum frequency of the arches below 10~Hz. Moreover, a control technique taking into account the mirror-bench differential signals was implemented in order to reduce their relative motion (BENCH-MIRROR), which is the quantity effectively responsible of the noise coupling. 

During O3, a malfunctioning was identified in the mechanical setting of the West Bench suspension (SWEB) which caused its actual motion to be comparable to the ground motion at the frequency of the main microseismic peak. Figure~\ref{fig:spectrograms} shows the mirror contribution and the bench contribution to the arches separately, for both North and West cavity, in two selected bad-weather conditions. In the West arm power spectrogram, the typical pattern is visible: the arches were entirely due to SWEB motion, and all the times the ground motion exceeded a certain threshold during the run, these arches entered the detector band. In the North arm power spectrogram, the arches were normally much lower, and the contribution from the bench motion was of the same magnitude as the mirror motion. It was even possible to find some special conditions (the largest component of the ground motion centered at 0.2~Hz), in which the mirror motion was prevalent (see Figure~\ref{fig:spectrograms}, bottom panel).

The issue concerning SWEB mechanics and control has been understood and cured after O3. In O4, its residual motion is expected to be at least similar to the one observed in O3 for SNEB. Further improvements in the control strategy will be tested for both the mirror and the bench suspension.

\subsubsection{Identification of scattered light culprits}

\begin{figure}[!htbp]
\centering
\begin{minipage}[c]{\textwidth}
\centering
\includegraphics[scale=0.49]{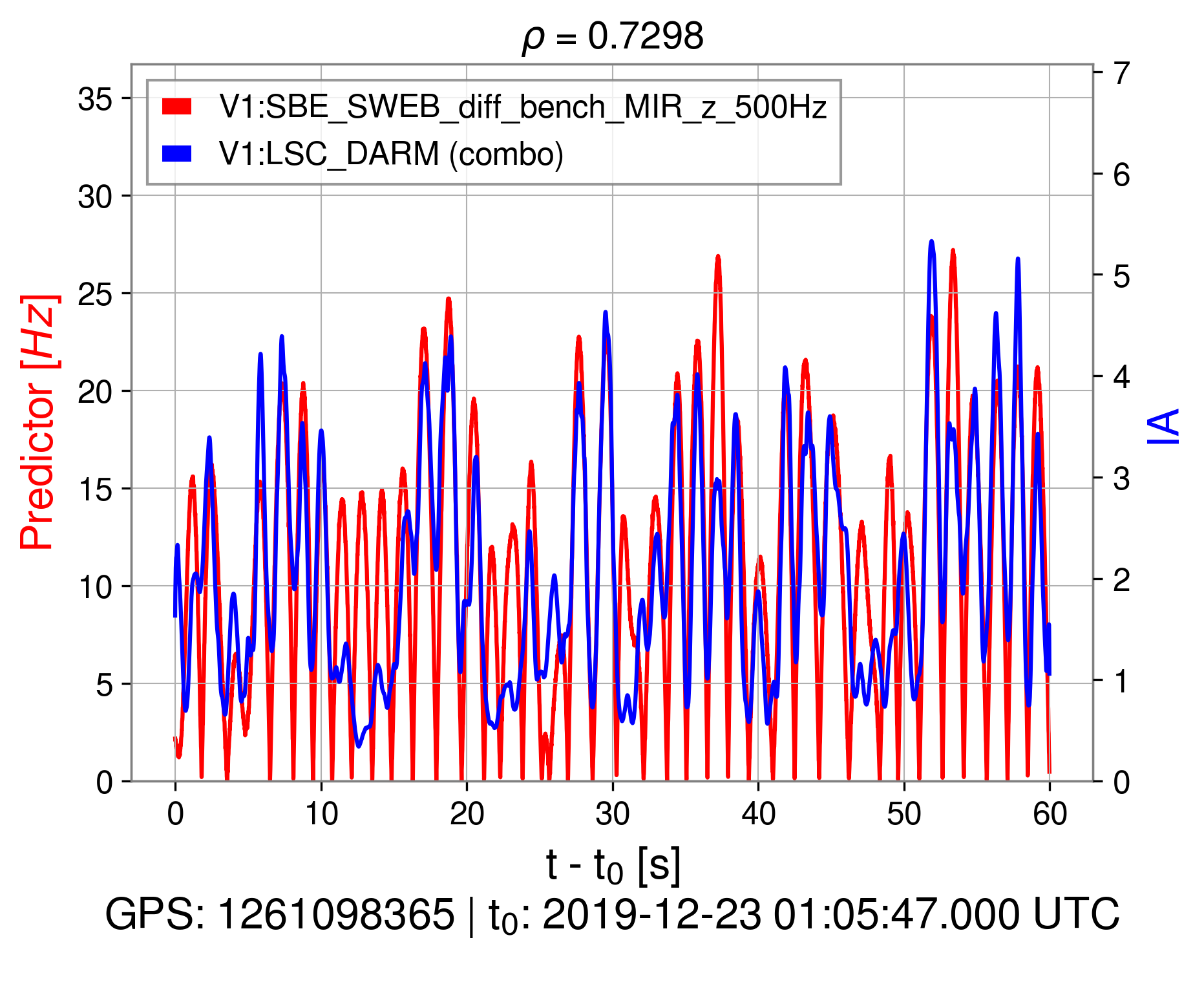}
\includegraphics[scale=0.49]{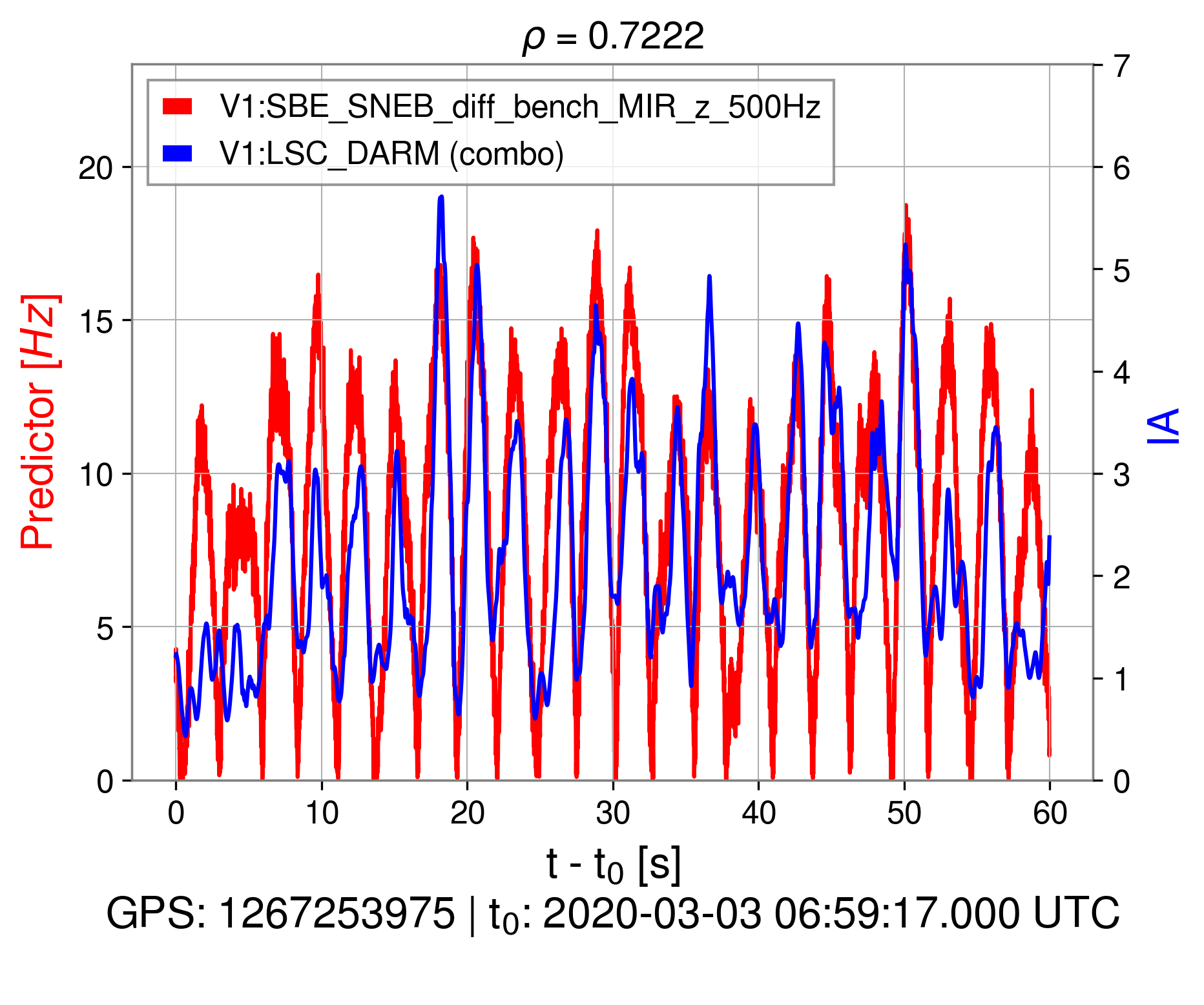}
    \caption{In red is the culprit's predictor, i.e. Equation~\ref{eq:arches} for the relative motion (diff) between the suspended end bench and the end mirror (BENCH-MIRROR) of the West end (left) and North end (right). The sum of the IA of the first two modes of DARM, extracted by tvf-EMD is shown in blue.}
    \label{fig:predictor}
\end{minipage}
\end{figure}

Part of the effort regarding SL noise mitigation consists in the localisation of SL sources, referred to as culprit, through data analysis. This can be a difficult and time consuming operation in a km-long detector with many possible sources of SL. Adaptive algorithms for time series analysis can be used to this end, due to their ability to decompose non-linear and non-stationary data into a set of oscillatory modes~\cite{Valdes:2017xce,Longo:2020onu,Longo:2021avq}. 
The methodology described in~\cite{Longo:2020onu} and based on the time varying filter empirical mode decomposition (tvf-EMD)~\cite{Li_2017} adaptive algorithm is applied to the two data segments shown in Fig.~\ref{fig:spectrograms}. SL noise couples with the differential motion of the arm cavities ({\em DARM}, the Virgo longitudinal degree of freedom sensitive to GW) time series, which is first low-passed and then decomposed using tvf-EMD to extract its oscillatory modes, from which the instantaneous amplitude (IA) is obtained using the Hilbert transform. Computing Equation~\ref{eq:arches} for a broad list of position sensors and correlating with the IA of DARM's oscillatory modes allows to quickly identify the most correlated channel, i.e. the culprit. The two data segment considered are
\begin{itemize}
\item GPS: 1261098365 UTC - 2019/12/23 01:05:47 + 60s,
\item GPS: 1267253975 UTC - 2020/03/03 06:59:17 + 60s. 
\end{itemize}
Obtained results are reported in Fig.~\ref{fig:predictor}, showing the predictors of the culprit for the end benches, based on Equation~\ref{eq:arches}, correlated with the IA of DARM. The culprits are related to the BENCH-MIRROR channel in both cases. The resulting values of correlation are $\rho=0.73$ for SWEB and $\rho=0.72$ for SNEB. Since after low-passing the data the first two oscillatory modes of DARM were found to be the most correlated with the same predictor, the sum of their IA is considered and is shown in Fig.~\ref{fig:predictor} for both cases, referred to as \emph{combo}. As a counter proof, in Fig.~\ref{fig:spectrograms} the predictors of the culprits are overlapped on the spectrograms of the WEB and NEB photodiodes. It can be seen that they closely match the SL arches. In particular, for the SWEB case, the mirror motion is small and the bench motion is mainly responsible for the observed SL. For SNEB case, while the mirror motion is significant the BENCH-MIRROR predictor, identified with adaptive analysis, better matches the arches also in this case. 

\subsection{Impact of wind}

\begin{figure}[!htbp]
    \begin{center}
      \includegraphics[width=0.98\columnwidth]{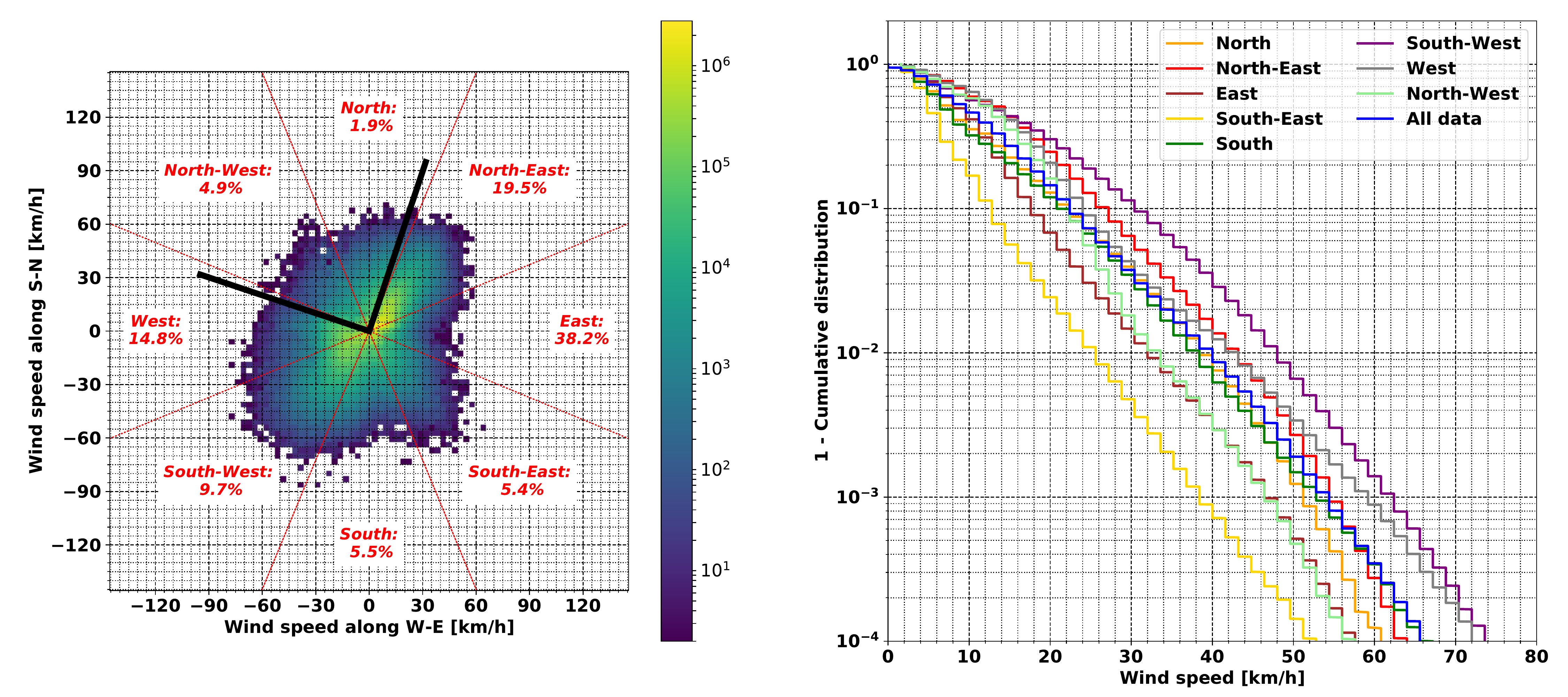}
    \end{center}
    \caption{Wind statistics as measured by the EGO weather station during the O3 run. The left plot shows the joint distribution of the wind speed and orientation, with the two black bars showing the directions of the two arms of the Virgo detector. The right plot shows the complementary cumulative distribution of the wind speed for each of the eight quadrants of the wind rose.}
    \label{fig:wind_2_O3_wholeRun}
\end{figure}

\begin{figure}[!htbp]
    \begin{center}
      \includegraphics[width=0.98\columnwidth]{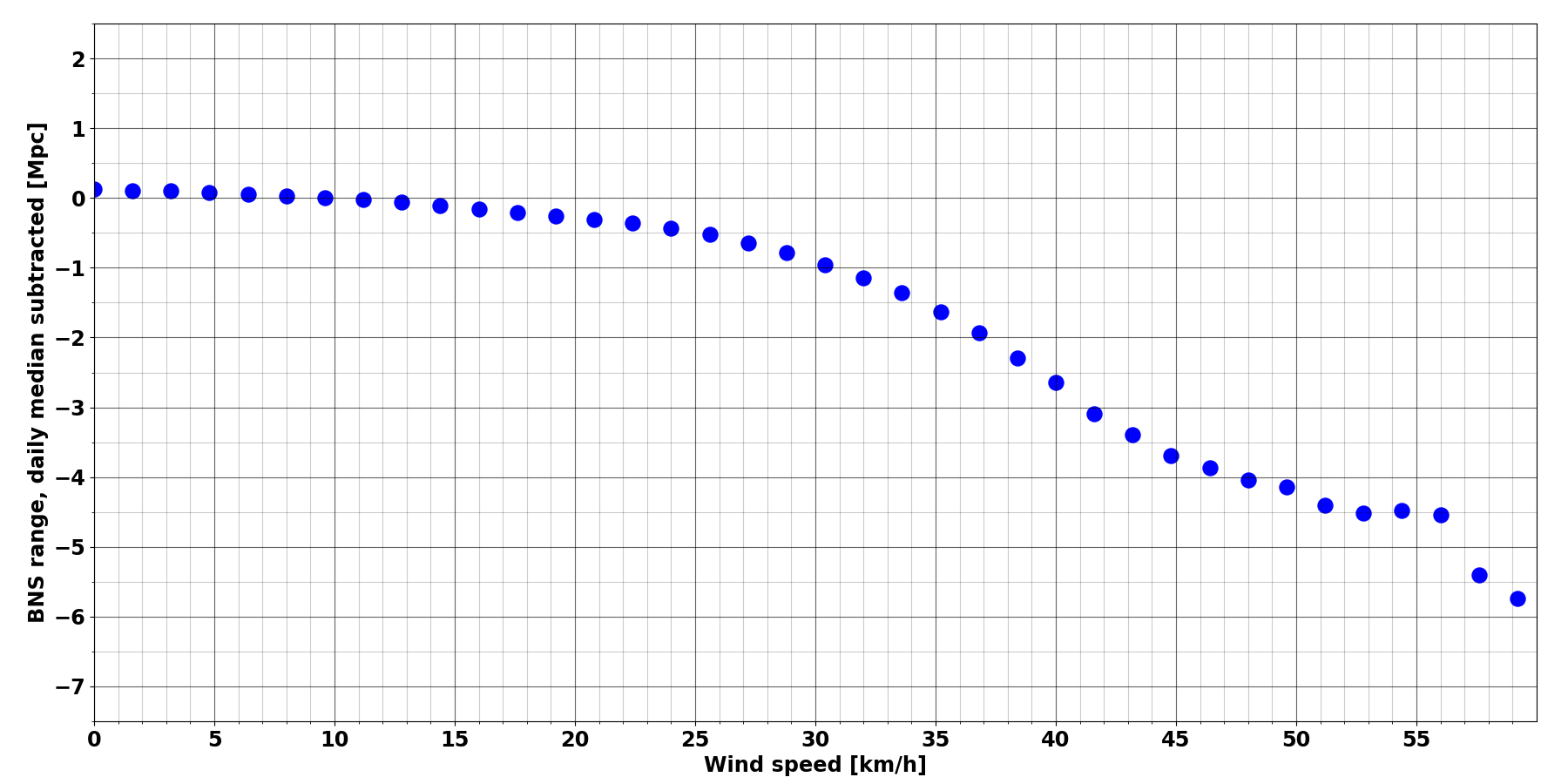}
    \end{center}
    \caption{Average variation of the BNS range around its local average, as a function of the wind speed. In the Virgo DAQ, the BNS range and the wind speed are updated every 4 and 2 seconds, respectively.}
    \label{fig:BNS_range_wind_variations}
\end{figure}

Figure~\ref{fig:wind_2_O3_wholeRun} summarizes the wind statistics recorded at EGO during the O3 run. Wind is blowing more often from the East while the stronger winds are predominantly coming from the West~-- the sea shore.
The method described in Sec.~\ref{section:impact_sensitivity} can be applied to quantify the impact of the instantaneous wind speed on the sensitivity. Figure~\ref{fig:BNS_range_wind_variations} shows that the sensitivity is pretty much unaffected until a wind speed of ${\sim} 20-25$~km/h, while the detector gets sensitive to larger speeds: the BNS range decrease exceeds ${\sim} 4$~Mpc for a wind speed of 50~km/h or above.
Yet this variation is limited (about 10\% of nominal BNS range values during O3), meaning that the detector is quite robust against wind. Another consequence of high-wind conditions is the need for the Virgo global control system to use larger corrections to keep the instrument at its nominal working point. And the larger these corrections, the more the detector is vulnerable to additional disturbances that could make the corrections saturate and lead to an almost immediate control loss.

\begin{figure}[!htbp]
    \begin{center}
      \includegraphics[width=0.98\columnwidth]{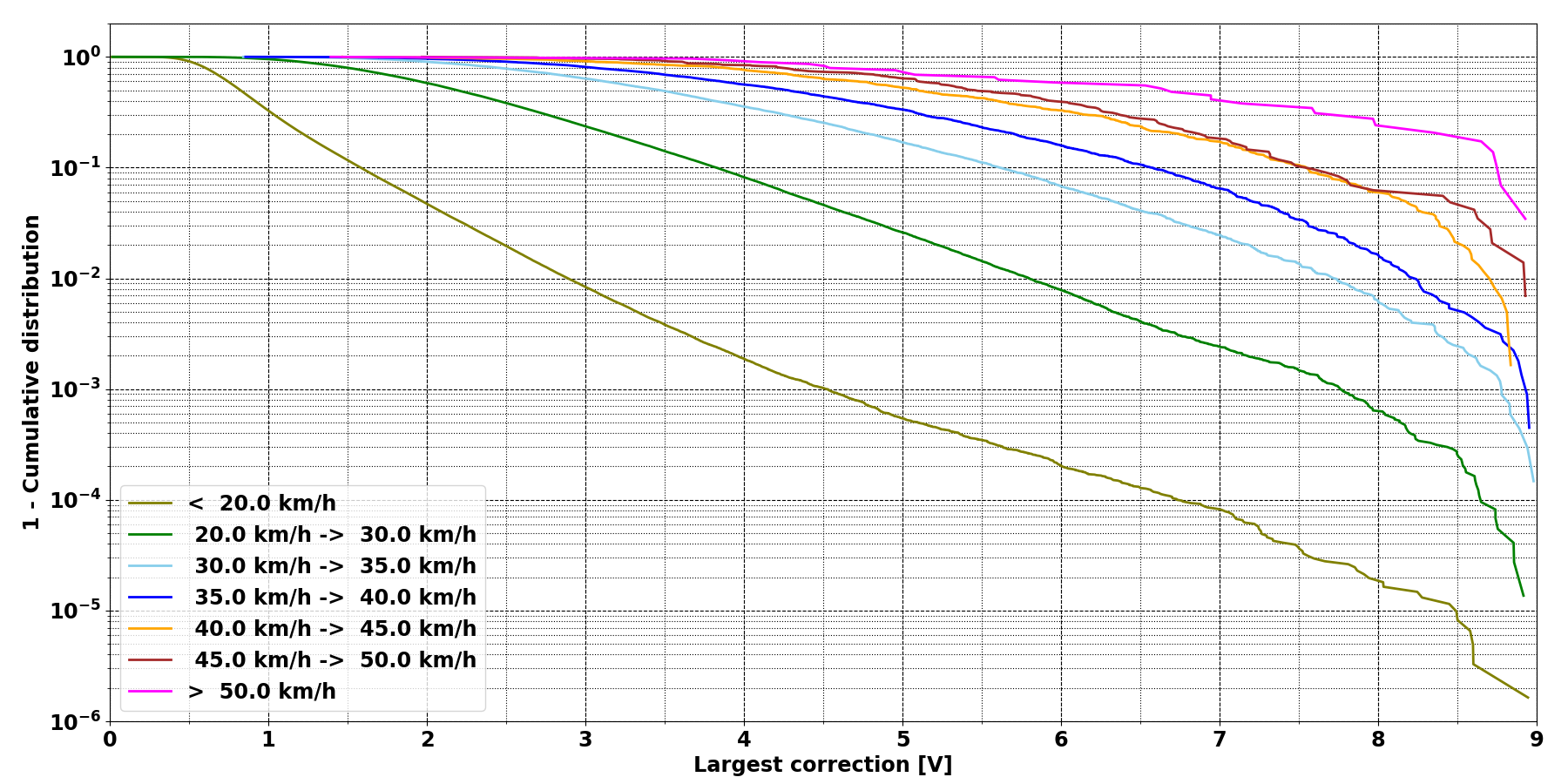}
    \end{center}
    \caption{O3 complementary cumulative distribution functions of the maximum longitudinal corrections (in volts) keeping the Virgo arm cavities resonant for different wind speed ranges. The mean wind speed and the maximal corrections have been computed over 30~s time windows. The $x$-axis ends at 9~V, a bit below the saturation level of 9.5~V for that particular correction.}
    \label{fig:max_DARM_correction_and_wind}
\end{figure}

The effect of the wind speed is clearly visible on Fig.~\ref{fig:max_DARM_correction_and_wind} that compares the complementary cumulative distribution functions of the kilometric Fabry-Perot cavity longitudinal corrections for different ranges of wind speed. Clearly, the larger the wind speed, the higher the correction. On this plot, the average wind speed and the maximum correction have been computed using non-overlapping time windows of 30~seconds each. 
The largest displayed correction range stops on purpose at 9~V because the actual physical correction saturates at 9.5~V, a value that can be reached or even exceeded when there is a control loss.
As the control system has some small but non-zero internal latency, it is not always clear whether the observed saturation is the cause of the control loss or a consequence of it. Therefore, for a cumulative plot like the one shown on Fig.~\ref{fig:max_DARM_correction_and_wind}, corrections above 9~V have been cut away to avoid contamination from correction signals posterior to control losses.

\subsection{Disentangling sea activity and wind}

\begin{figure}[!htbp]
    \begin{center}
      \includegraphics[width=0.98\columnwidth]{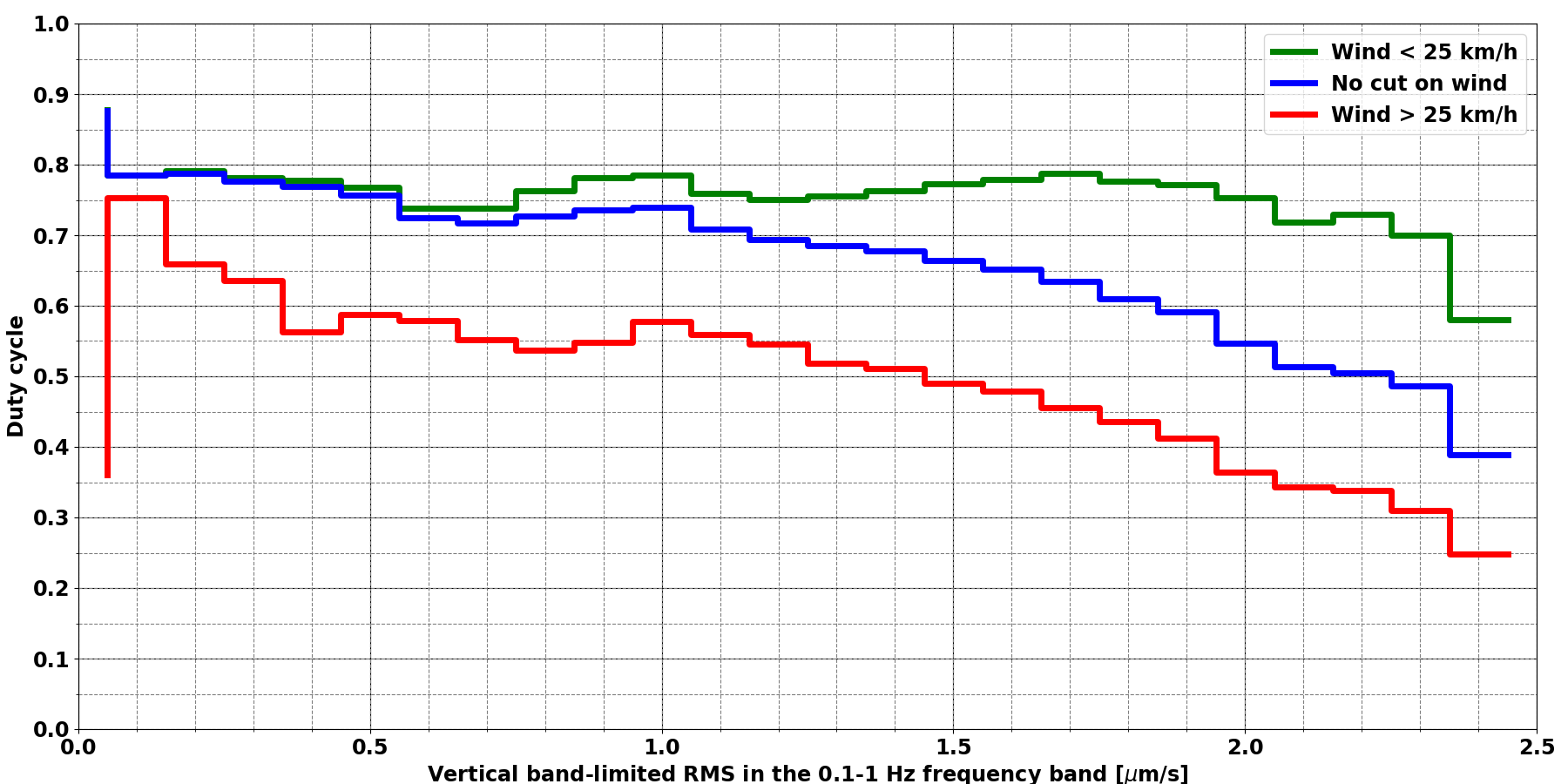}
    \end{center}
    \caption{Virgo duty cycle during the O3 run versus microseism activity, for three different wind conditions: blue~$\leftrightarrow$~no cut on wind speed; green~$\leftrightarrow$~low wind (speed below 25~km/h); red~$\leftrightarrow$~high wind (speed above 25~km/h).}
    \label{fig:duty_cycle_sea_activity_wind}
\end{figure}

Fig.~\ref{fig:duty_cycle_sea_activity_wind} attempts to disentangle the impact of high microseism levels (due to the nearby rough sea) and high wind,
by looking at the O3 Virgo duty cycle as a function of the microseism level for three different wind conditions: no cut on wind speed (blue histogram); low wind speed (below 25~km/h, green); high wind speed (above 25~km/h, red). One can see that in low wind conditions the duty cycle is pretty much independent from microseismicity, whereas it is lower and decreases more quickly when the wind level increases. Therefore, the Virgo detector appears robust against microseism but more sensitive to wind. Note that the extreme bins on the histograms plotted on Fig.~\ref{fig:duty_cycle_sea_activity_wind} may have low statistics compared to others (low wind and high microseism, or high wind and low microseism are rare conditions): this explains why the duty cycles reported there fluctuate significantly compared to neighboring bins.

\section{Other environment impacts}
\label{section:other}
\markboth{\thesection. \Sectionname}{}
Additional sources of external noise have potential impact on the interferometer. Hereafter we describe those sources that we have further investigated during O3, namely: Schumann's resonance magnetic fields, lightning strikes and cosmic ray muons.

\subsection{Magnetic noise}

\begin{figure}[!htbp]
\centering
{\includegraphics[width=0.98\textwidth]{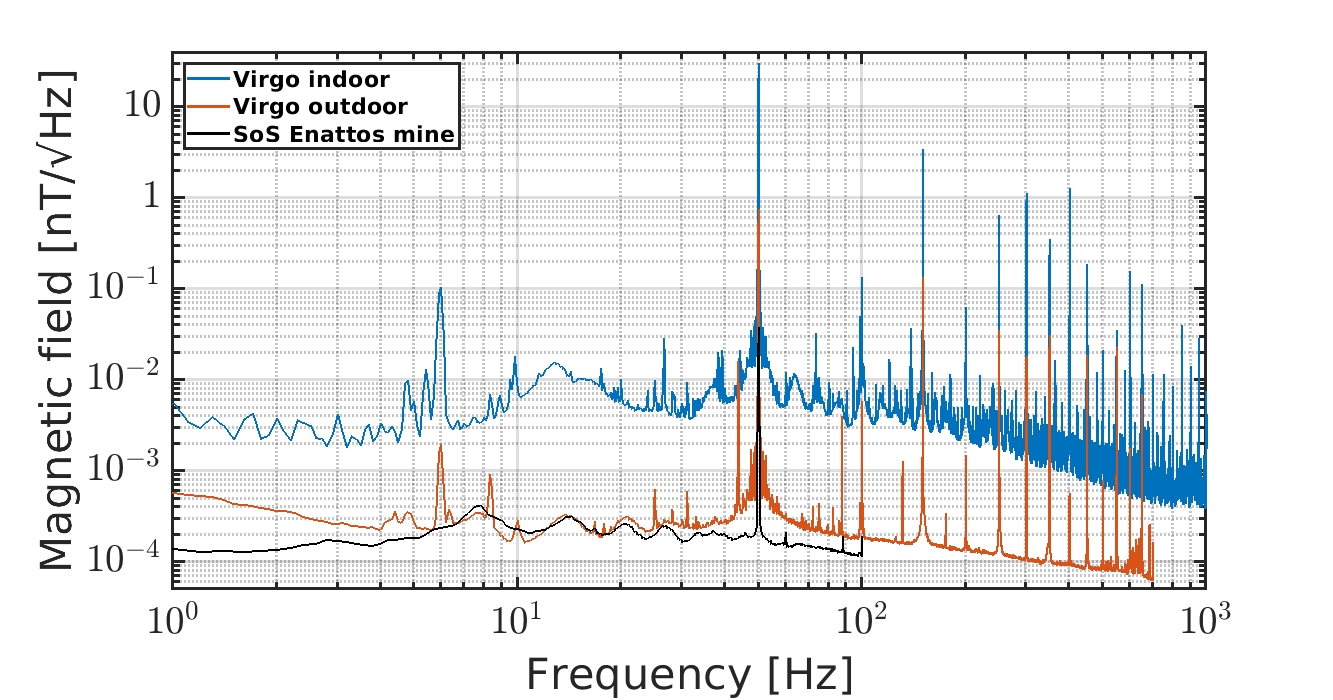}} 
\caption{Amplitude spectral densities of indoor (blue curve) and outdoor (red curve) magnetometers at EGO and at Sos Enattos mine in Sardinia (black curve). 
The quiet Sos Enattos location shows evidence of Schumann resonances peaked at approximately 8, 14, 21, 27 and 33~Hz.}
\label{fig:magnetic_spectra}
\end{figure}

Ambient magnetic fields can couple to GW interferometers, for example through the magnetic actuators used for the control of the seismic isolation platforms of optical components and of the test masses~\cite{Virgo_env_O3,Cirone_mag_pay}.
Like gravitational waves, electromagnetic (EM) waves travel at the speed of light, and, due to their strength, could affect multiple detectors with time differences compatible with those expected from some GW.

Magnetic fields that extend over the entire planet, such as the Schumann resonances~\cite{Schumann} (SR), or large-current lightning strikes, can limit the sensitivity to GW signals correlated over multiple detectors~\cite{StochSchumann, Kowalska_Leszczynska_2017}.
One purpose of the EGO external magnetometers (see Sec.~\ref{section:virgo}) is to monitor the level of these global magnetic fields.

At Virgo, the external magnetic environment is much quieter than inside experimental halls where stray magnetic fields are radiated by electric loads and cables where large currents are circulating. Figure~\ref{fig:magnetic_spectra} compares inside and outside magnetometer spectra recorded at Virgo during O3 and in the very quiet environment inside the Sos Enattos mine in Sardinia~\cite{Naticchioni_2020}.
The most intense spectral noise features are narrow lines at the 50~Hz electric mains frequency and its odd harmonics. The RMS amplitude of the 50~Hz line measured at Virgo is of the order of 0.1~nT in the external location, while it is at least 50 times larger in any inside location.

\begin{figure}[!htbp]
    \begin{center}
\includegraphics[width=0.9\columnwidth]{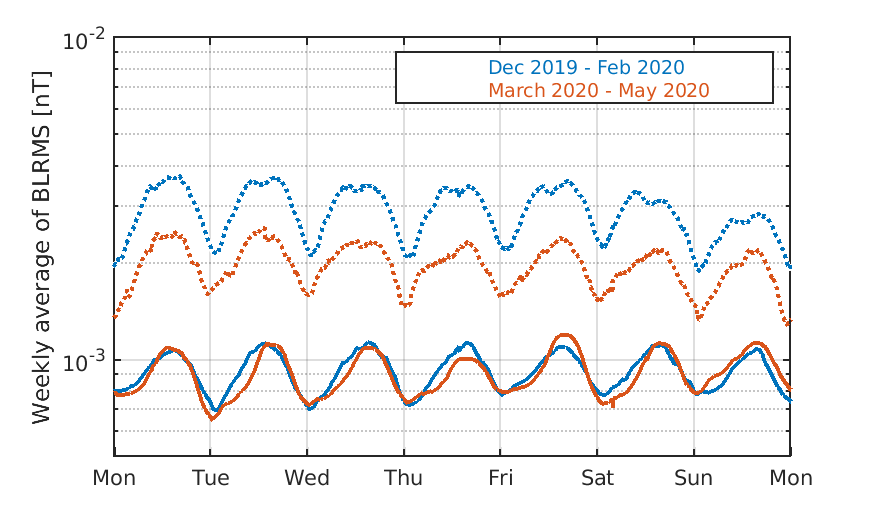}
   \end{center}
    \caption{
Weekly averaged magnetic field band-limited RMS values computed in two frequency bands: 
1 to 6~Hz (dashed) and 18 to 24~Hz (solid). 
Magnetic field intensity is measured externally of Virgo experimental buildings, in the reference period between December 2019 and February 2020 (blue curves) and in the period between March 15 and May 15 (red curves) which corresponds to reduced anthropogenic activity within and outside of EGO because of the COVID-19 pandemics.
}
    \label{fig:magnetic_BRMS}
\end{figure}

Virgo external magnetometers detect the SR field.
This consists of steady EM waves that resonate inside the waveguide formed by the Earth surface and the ionosphere, and which are excited by globe-wide lightning activity. 
The second and third SR modes (peak frequency around 14~Hz and 21~Hz, respectively) are visible above noise at almost any time, their median amplitude during O3 is a few tenth of pT, their intensity follows a 24-hour modulation.
The measured daily modulation of the third SR mode is shown in Fig.~\ref{fig:magnetic_BRMS}.  This modulation is thought to be associated to temperature-driven variations in the height of the ionosphere EM waveguide~\cite{Volland}.
The first SR mode and those of order greater than three, are often covered by anthropogenic magnetic noise. Figure~\ref{fig:magnetic_BRMS} shows that during the COVID-19 lockdown period from March to May 2020,
the external magnetic field median RMS in the low frequency region from 1 to 6~Hz reduced by about 50$\%$ with respect to the reference period between December 2019 and February 2020.
At the same time, the magnetic field RMS amplitude between 18~Hz and 24~Hz around the $3^{rd}$ Schumann mode, did not change appreciably.

At EGO, anthropogenic external magnetic noise follows a daily modulation: broad maxima during working hours and minima around 01:00 LT. This noise has the form of short transients with intensity of $ \approx 10$~pT extending from DC up to approximately 20~Hz. We believe this noise is associated to train transits along railway tracks at about 6~km distance from the site. The sudden trunk-line change when a train passes from an electro-duct section to another one creates stray currents and magnetic fields that are observed as magnetic glitches at EGO.
According to the measured coupling of ambient fields~\cite{Virgo_env_O3}
we estimate a negligible impact of Schumann's and anthropogenic magnetic noise on the sensitivity of the future Virgo upgrades. More relevant might be the impact of the correlated Schumann noise on multiple interferometers, which is under evaluation.

\subsection{Lightnings}

Lightning strikes produce prompt EM waves and much slower air pressure waves which induce vibrations of the ground and of the detector mechanical components.
There are studies of correlated lightnings noise between the Virgo and LIGO sites~\cite{Kowalska_Leszczynska_2017} and at the KAGRA underground observatory~\cite{Washimi_2021b}.

\begin{figure}[!htbp]
    \begin{center}
      \includegraphics[width=0.98\columnwidth]{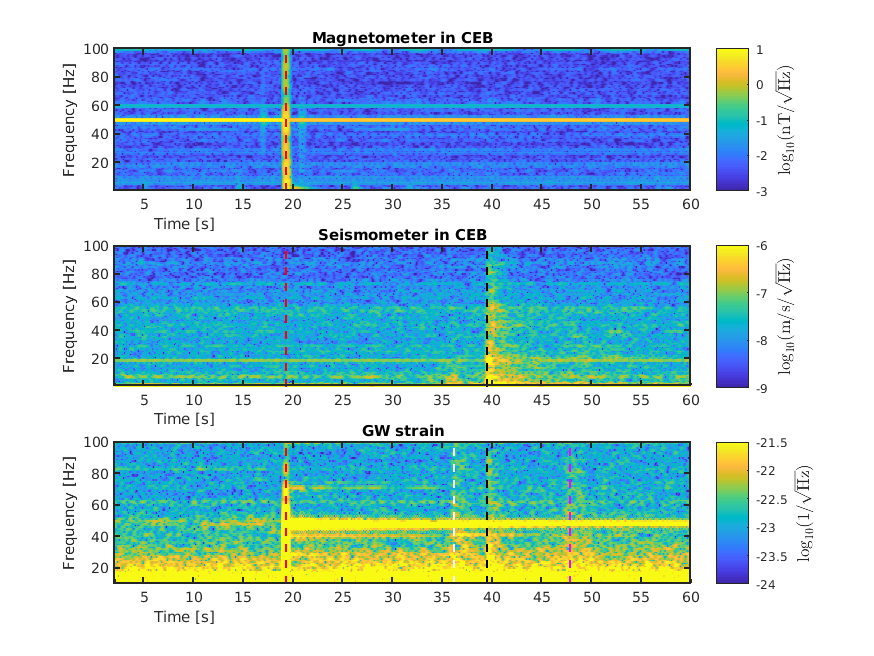}
   \end{center}
    \caption{Impact on the Virgo environment and detector of one lightning strike which occurred 6 to 10~km away from Virgo buidings on November 15, 2019 at 23:25:51 UTC. The spectrograms of a few relevant signals are shown.
(Top) A prompt magnetic transient is detected by magnetometers at the time of the event, marked by the red vertical line. (Middle) A few tens of seconds later, a seismic (and acoustic, not shown) transient is detected in the central experimental area, marked by the black vertical line. 
The bottom spectrogram shows the reconstructed GW strain during the same time interval. 
The red vertical line marks the lightning strike occurence, the black, magenta and white vertical lines mark the occurrence of seismic transients detected in the
CEB, NEB and WEB, respectively.
}
    \label{fig:lightning}
\end{figure}

The typical effect of the impact of a lightning strike occurring at approximately 10 kilometers from the Virgo detector during O3 is illustrated in Fig.~\ref{fig:lightning}.
A distinctive feature of lightning strikes is a coincident short transient noise in magnetometers located inside the 3~km-distant Virgo experimental buildings (top graph of Fig.~\ref{fig:lightning}).  
The magnetic impulse is followed by the slower sound shock wave detected by seismometers (middle graph of Fig.~\ref{fig:lightning}). The bulk of displacement noise reaching the buildings is below 10~Hz.

The bottom graph of Fig.~\ref{fig:lightning} illustrates the effect of the lightning in the GW strain signal.
In coincidence with the spike in magnetometers, we observe a prompt broadband low-frequency noise and the onset of a 48~Hz narrow spectral noise, with a minute-long decay time, leading to a ${\sim} 30$\% drop of the live BNS range.  
This latter noise has been associated to one structural mode of the West end test mass suspension, which gets excited because of the coupling of ambient magnetic fields with the magnetic actuators located along the suspension.
Moreover, associated with the delayed acoustic and seismic bursts of ambient noise reaching the experimental buildings, a broadband strain noise shows up, extending up to about 100~Hz. This is likely due to scattered light processes within the interferometer.

Data quality flags triggered by lightning strikes were produced during the O3 run; they proved useful in a test aiming at filtering out part of the false-alarm triggers found by a real-time transient GW search~\cite{o3virgodetchar}. Further studies are planned during the O4 run preparation. 

\subsection{Cosmic muons}\label{subsec:muons}

Ground-based GW detectors are constantly passed through by \emph{muons}, produced by the interaction of cosmic rays with Earth's atmosphere~\cite{maurin2014database}. 
These energetic charged particles can interact with the detector test masses and constitute an additional source of noise, as addressed in the literature since the first prototypes of resonant mass GW detectors~\cite{PhysRevLett.23.184,amaldi1986estimate,GIAZOTTO1988241,CHIANG1992603}.

\begin{figure}[!htbp]
	\begin{center}
		\includegraphics[width=.8\textwidth]{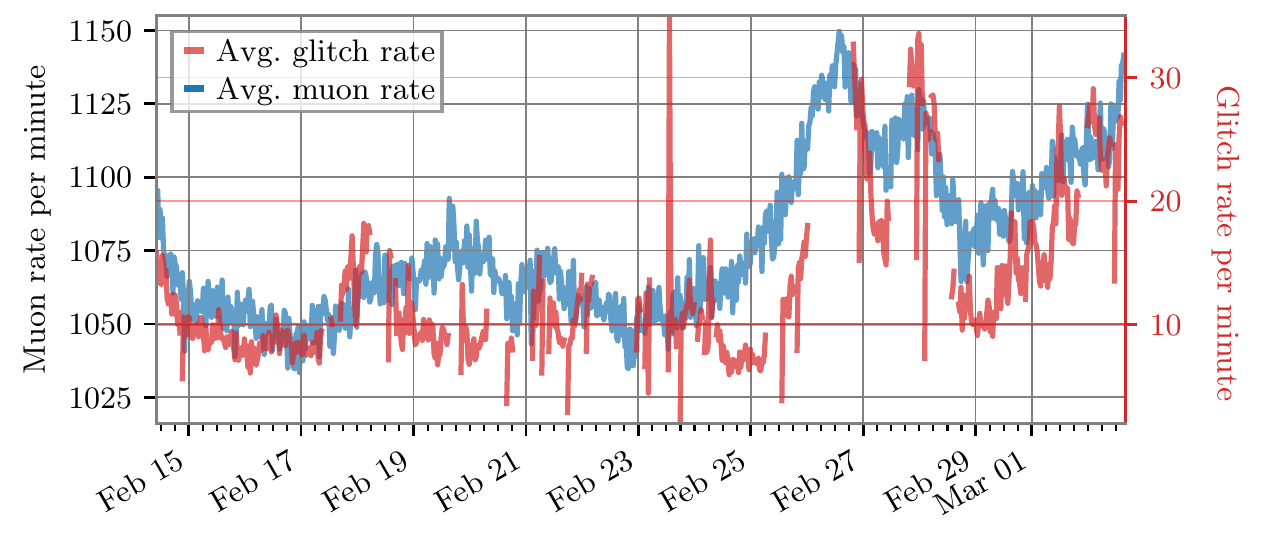}
	\end{center}
	\caption{Correlation between muon and glitch rates. The blue line represents the time series of the average rate per minute of muons while the red one is the time series of the rate per minute of glitches with $\mathrm{SNR}>4.5$ and frequency at peak in $[10,4096]~\mathrm{Hz}$ as identified by the Omicron pipeline~\cite{ROBINET2020100620}. Gaps in the latter correspond to periods when the detector was not taking data.}
	\label{fig:MuonVsGlitch}
\end{figure}

We report here the preliminary results on the first measurement of potential effects of these muons on the Virgo detector noise.
This study has been carried out by means of about $17$ days, at the end of the O3b run, of joint data acquisition of Virgo and a muon telescope designed by the IP2I laboratory~\cite{gi-1-33-2012}, installed in the CEB close to the beam splitter mirror.
Two kind of tests have been performed. 
In the first one, we have evaluated whether the rate of muons in the correspondence of GW candidate events was larger than the reference values of the period: we have found no statistical evidence of an excess of muons in correspondence of these triggers. In the second test, we have estimated the correlation of this rate with the rate of glitches in Virgo noise.
Figure~\ref{fig:MuonVsGlitch} shows the time series corresponding to the rates of glitches and muons, averaged on strides of 30 minutes. 
Here, a correlation is clearly evident.
This is actually not surprising, for the number of the muons arriving at ground being highly dependent on air density and ultimately on parameters like atmospheric pressure and temperature.
These quantities are also witnesses of the weather conditions, which in turn can determine an increase of the detector noise, as we have commented in Sec.~\ref{sec:seaimpact}.
Therefore, both the variations of these rates share the same main cause, which explain their large correlation.
Once the effects of the atmospheric conditions are removed via a regression analysis, the residuals exhibit no significant correlation.

\section{Outlook and prospects for O4}
\label{section:outlook}
\markboth{\thesection. \Sectionname}{}
The Virgo detector performances are affected by external environment conditions; in particular, seismic noise, earthquakes, bad weather, magnetic noise and lightnings have an impact on the detector sensitivity or duty cycle. The main coupling mechanisms are: direct excitation of suspended mirrors, vibration of experimental buildings, shaking of benches hosting auxiliary optical systems, disturbances on critical electronic equipment, SL.

If the detector control system is able to manage the effect of a disturbance, the interferometer can remain at its working point with a reduced sensitivity. Otherwise the global control gets lost and the procedure to recover it has to be started again from the beginning, thus impacting on the duty cycle~-- see~\ref{section:lock_losses}.

In this work we reported the results of the analysis of such events during the O3 run. Thanks to the large amount of data collected, we were able to perform a careful statistical analysis of the impact of several kinds of external disturbances.

The results confirm that the Virgo detector is a very robust apparatus. The 
sensitivity reduction due to anthropogenic seismic noise is very low: less than 2\% in terms of BNS range. Also the degradation due to the wind is limited: it appears only for wind speeds larger than 25 km/h, reaching a sensitivity reduction as large as 10\% only for very high speed (larger than 50 km/h).

In these cases, the sensitivity reduction is due to an increased noise at low frequency as well as to the appearance of short high frequency glitches.
In few cases, such kind of noise was indirectly originated by lightnings.

Both microseism and wind have an impact on the detector duty cycle, since the increasing correction signals acting on the mirror during bad weather can saturate, finally resulting to a control loss. It results that the Virgo detector global control is more robust against microseism while it is less effective against strong wind.

The analysis of control losses during O3 confirms that earthquakes are a relevant source of these. The Seismon framework, useful to keep the detector in a safe state to try to avoid loosing its control during such events, was used during the whole O3 run and it is now being upgraded for the next scientific run.

An upgrade of the EMS is in progress to better face the influence of external disturbances: installation of a new lightning detector in the central area; installation of two additional weather stations at the end buildings to monitor local wind gusts; and the installation of more sensitive accelerometers on locations prone to light scattering (viewports, external optical benches, etc.).

These actions, together with several other upgrades of the Virgo detector, already performed or presently in progress, will have a crucial role for the success of the next scientific run O4, which is expected to start at the end of 2022.

\section*{Acknowledgements}
\markboth{\Sectionname}{}
The authors gratefully acknowledge the Italian Istituto Nazionale di Fisica Nucleare (INFN),  
the French Centre National de la Recherche Scientifique (CNRS) and
the Netherlands Organization for Scientific Research (NWO), 
for the construction and operation of the Virgo detector
and the creation and support of the EGO consortium.
The authors also gratefully acknowledge research support from these agencies as well as by 
the Spanish  Agencia Estatal de Investigaci\'on, 
the Consellera d'Innovaci\'o, Universitats, Ci\`encia i Societat Digital de la Generalitat Valenciana and
the CERCA Programme Generalitat de Catalunya, Spain,
the National Science Centre of Poland and the European Union – European Regional Development Fund; Foundation for Polish Science (FNP),
the Hungarian Scientific Research Fund (OTKA),
the French Lyon Institute of Origins (LIO),
the Belgian Fonds de la Recherche Scientifique (FRS-FNRS), 
Actions de Recherche Concertées (ARC) and
Fonds Wetenschappelijk Onderzoek – Vlaanderen (FWO), Belgium,
the European Commission.
The authors gratefully acknowledge the support of the NSF, STFC, INFN, CNRS and Nikhef for provision of computational resources.
	
{\it We would like to thank all of the essential workers who put their health at risk during the COVID-19 pandemic, without whom we would not have been able to complete this work.}

\appendix
\addtocontents{toc}{\addtolength{\cftsecnumwidth}{5.5em}}
\addtocontents{toc}{\addtolength{\cftsubsecnumwidth}{5em}}

\section{Study of the control losses during O3}
\label{section:lock_losses}
\markboth{\uppercase{\thesection.} \Sectionname}{}

The Virgo detector needs to be controlled accurately in order to be sensitive to gravitational-wave signals~\cite{ACERNESE2020102386,galaxies8040085}. Schematically, there is an automated procedure~\cite{o3virgodetchar}  that brings the instrument from an initial state where the optics and the laser are controlled independently one from another, to the nominal state where the different optical cavities are jointly resonant and the interferometer itself is used as a length etalon to control further the laser frequency. That procedure typically takes about 15-20 minutes and requires 1-2 attempts to complete. Then, the global control of the detector is kept as long as possible, with feedback loops maintaining Virgo at its nominal working point. When that control is lost for whatever reason, data taking stops and the control acquisition procedure has to be started again. This leads to a decrease of the instrument duty cycle and can cause transient gravitational waves to be missed. Therefore, it is important to find out the causes of the control losses and to use this information to improve the feedback systems and make them more robust.

As explained in Sec.~\ref{section:earthquakes} above, a global study of the control losses was needed to be able to extract those likely due to earthquakes. It was decided to focus on the 601 control losses that occurred during O3 while the detector was taking data in nominal conditions (Science mode), to be sure that no particular human action was happening on the instrument at any of these times. Related to the duration of the O3 run (about 11 months) and to the duty cycle of the Virgo detector (about 75\%), this corresponds to about 1 control loss every 10 hours of data taking on average. And, in reality, uninterrupted data taking stretches could be much longer as control losses usually cluster in time when a particular problem impacts the detector.

The first part of the study was to define the time the control loss occurred for each of these events. For that, we have used three different Virgo DAQ channels.
\begin{itemize}
\item Two fast channels, sampled at 10~kHz: \texttt{ARM\_POWER}, latching when the power stored in the kilometric arm cavities goes below some threshold, meaning that they are not resonant anymore; \texttt{DARK\_FRINGE\_SHUTTER}, triggered when the fast shutter protecting the dark fringe photodiodes from an excess of light~\cite{TheVirgo:2014hva} closes.
\item One slow channel, sampled at 1~Hz: \texttt{AUTOMATION\_STATUS}, monitoring the global status of the detector, as seen by the automation process that steers the instrument.
\end{itemize}

\begin{table}[!htbp]
\begin{center}
\begin{tabular}{|c|c|c|c|}
\hline \texttt{ARM\_POWER} & \texttt{DARK\_FRINGE\_SHUTTER} & \texttt{AUTOMATION\_STATUS} & Total \\
\hline                  14 &                            559 &                          28 &   601 \\
\hline
\end{tabular}
\caption{Number of control losses in Science mode  witnessed first by each DAQ channel used to time accurately control losses. As expected, the two fast channels are by far those that detect a control loss first. Most of the time the fast shutter protecting the dark fringe photodiodes closes before the arm power loss gets large enough to trigger the other fast channel.}
\label{table:lock_loss_witnesses}
\end{center}
\end{table}

The time of a control loss is defined as the earliest time one of these three switches flips from its nominal value to the value corresponding to an uncontrolled detector. Most of the time, as expected, the fast channels are the first ones to latch. And they do almost simultaneously, given that the cavity resonance losses are all connected. Though, in practice, the dark fringe shutter closes almost always before the cavity arm power has decreased below its nominal threshold. In addition there are a few cases for which the central automation system triggers first a shutdown of the detector global control, either because it has detected an issue or because it has received a manual abort request from the operator on duty. Table~\ref{table:lock_loss_witnesses} shows the breakout of witnesses for the O3 control losses that occurred while taking Science data.

Then, the selected strategy consists in testing several hypothesis in parallel for each of these events~-- the main hypothesis investigated are listed in Tabs.~\ref{table:sure_lock_losses} and~\ref{table:other_lock_losses} and documented in the neighbouring text.

Various algorithms scanning the data around the control loss have thus been developed, with the twofold goals of being
\begin{itemize}
\item {\it complete}: to have as many control losses as possible tagged by at least one control loss hypothesis; 
\item {\it selective}: to find the right control loss origin as often as possible.
\end{itemize}
Achieving (close to) completeness requires testing many hypothesis, while a profusion of algorithms could be detrimental to the selectivity of the method. Therefore, the classification starts with a subset of hypothesis, those that, when identified, certainly cause a control loss and are also very likely to be the root cause of that particular event. Obvious examples in that category~-- called {\em sure} in the following~-- are control losses induced manually by the operator on duty, or hardware problems unambiguously identified by the real-time monitoring system of the Virgo detector. These control loss hypothesis are independent by definition and the associated algorithms should be selective. This has been checked by processing the 601 O3 control losses studied. All these events have been associated with at most one control loss hypothesis belonging to the sure category: 24\% with one, 76\% with none.

\begin{table}[!htbp]
\begin{center}
\begin{tabular}{|c|c|c|c|c|c|c|}
\hline Error & Manual & Hardware & Control software & PI & Earthquakes &        Total \\
\hline     2 &     10 &       92 &                7 &  2 &          30 &   143 (24\%)\\
\hline
\end{tabular}
\caption{Sure causes for 143 O3 control losses in Science mode~-- see text for details.}
\label{table:sure_lock_losses}
\end{center}
\end{table}

Table~\ref{table:sure_lock_losses} provides details about the 143 control losses whose cause has been tagged as sure, as described above. The dominant class is hardware problems, mainly transient interruptions of the data flow coming from some suspensions and causing feedback control systems to fail. The faulty components have been identified and replaced during the post-O3 shutdown and upgrade phase. Therefore, these problems are not expected to reoccur during the O4 run. Then, earthquakes are the second most common source of control losses in the sure category; about three times a month on average. Manual control losses induced by the operator on shift follow: they are due to the need to switch from nominal data taking to another task: weekly maintenance, regular calibration or commissioning activity. In O4 and beyond, such control losses should no longer occur as the procedure will be updated to require leaving Science mode before manually aborting the control. In 7 cases (only 1\% of the total control losses) the source of the event could be traced to some software problem; 2 more cases were due to human errors.

Finally, two control losses are labelled as {\em PI} for parametric instabilities, an optomechanical phenomenon due to the interaction between optical and mechanical modes of the detector and that had been observed at LIGO in 2015 before finally being seen in Virgo as well in January 2020~\cite{PI_O3b}. If not mitigated, a PI can make control systems saturate in a deterministic way (meaning that the saturation will consistently reoccur as long as the detector remains in a configuration favourable for its appearance and growth), thus impacting the detector duty cycle. Moreover, it is impossible to predict exactly what combinations of the instrument parameters will lead to a PI. Therefore, a dedicated simulation framework has been developed to estimate the susceptibility of Virgo to PIs during O3, for O4, and beyond~\cite{PI_CohenEtAl}.

\begin{table}[!htbp]
\footnotesize
\begin{center}
\begin{tabular}{|c|c|c|c|c|c|c|c|c|}
\hline \begin{tabular}{@{}c@{}}Fast \\ unlocks \end{tabular} & \begin{tabular}{@{}c@{}}Actuation \\ saturation \end{tabular} & \begin{tabular}{@{}c@{}}DARM \\ control \\ inaccuracy \end{tabular} & \begin{tabular}{@{}c@{}}Power \\ loss in \\ sidebands \end{tabular} & \begin{tabular}{@{}c@{}} Arm \\ power \\ asymmetry \end{tabular} & \begin{tabular}{@{}c@{}} Likely \\ missing \\ data \end{tabular} & \begin{tabular}{@{}c@{}} Automation \\ decision \end{tabular} & Others & Total \\
\hline 173 & 85 & 77 & 22 & 4 & 10 & 23 &\begin{tabular}{@{}c@{}} 64 \\ (11\%) \end{tabular} & \begin{tabular}{@{}c@{}} 458 \\ (76\%) \end{tabular} \\
\hline
\end{tabular}
\caption{
Breakout by category of control losses not tagged as sure.
64 (about 11\% of the total number of control losses recorded in Science mode during the O3 Virgo run) control losses have not been accurately classified, either because none of the tested hypothesis seemed to match the recorded data or because too many hypothesis were found matching, making their classification unconclusive. Further studies will be done when pre-O4 control losses data become available, in order to make the current classification more complete.}
\label{table:other_lock_losses}
\end{center}
\end{table}

Table~\ref{table:other_lock_losses} describes how the remaining control losses $({\sim} 76\%)$ have been classified. 11\% of the total remain unclassified, either because none of the hypothesis tested matched, or because too many did and there was no clear way to find out which one was the root cause (if identified).

The largest category by far (29\%) are the so-called {\em fast unlocks}, events that are almost instantaneous and occur within the laser injection system, upstream of the interferometer. Such control losses have been present for years, at rates that strongly vary over time, ranging from crisis periods lasting some hours to very quiet times.
This past Summer, following detailled investigations of the fast unlock characteristics, this problem has been finally solved by installing~\cite{logbook52813} a pigtailed Electro-Optic-Modulator (EOM) with a larger dynamic, and thus able to compensate the laser frequency glitches that were found to be the cause of fast unlocks. They have not reoccurred since this new EOM has been operated.

The next five categories are all related to the variety of feedback control systems that are running in parallel to keep the whole detector at its nominal working point. Improving the accuracy and the robustness of these systems while making the instrument more complex and thus more sensitive to the passing of gravitational wave is a permanent challenge, taken up during each upgrade or commissioning phase.

The analysis of the O3 control losses has been made using two independent software frameworks whose results have been compared: they have been found in good agreement, in particular for the dominant control loss categories. With the experience gained during O3, the goals for O4 are to improve the monitoring of the control losses and to reduce the latency of their analysis. A software framework similar to the Data Quality Reports (DQR)~\cite{o3virgodetchar,dqr,o3ligodetchar} used to vet in real time the gravitational-wave transient candidates that are significant enough to trigger a public alert is under development:
here control losses play the role of GW event candidates and the set of checks run to assess the data quality is replaced by the test of the various hypotheses for control loss.
This upgraded tool will be available to improve the overall performance of the instrument during the commissioning phase and associated noise-hunting for the new, double-recycled Advanced Virgo interferometer.

\section*{References}
\bibliographystyle{iopart-num}
\bibliography{references}

\providecommand{\newblock}{}
\begin{thebibliography}{10}
\expandafter\ifx\csname url\endcsname\relax
  \def\url#1{{\tt #1}}\fi
\expandafter\ifx\csname urlprefix\endcsname\relax\def\urlprefix{URL }\fi
\providecommand{\eprint}[2][]{\url{#2}}

\bibitem{TheLIGOScientific:2014jea}
Aasi J {\em et~al.\/} (LIGO Scientific) 2015 {\em Class. Quant. Grav.\/} {\bf
  32} 074001 (\textit{Preprint} \eprint{1411.4547})

\bibitem{TheVirgo:2014hva}
Acernese F {\em et~al.\/} (Virgo Collaboration) 2015 {\em Class. Quant.
  Grav.\/} {\bf 32} 024001 (\textit{Preprint} \eprint{1408.3978})

\bibitem{10.1093/ptep/ptab018}
Akutsu T {\em et~al.\/} 2021 {\em Progress of Theoretical and Experimental
  Physics\/} {\bf 2021} ISSN 2050-3911 05A102 (\textit{Preprint}
  \eprint{https://academic.oup.com/ptep/article-pdf/2021/5/05A102/38109702/ptab018.pdf})
  \urlprefix\url{https://doi.org/10.1093/ptep/ptab018}

\bibitem{Abbott:2016blz}
Abbott B {\em et~al.\/} (LIGO Scientific Collaboration, Virgo Collaboration)
  2016 {\em Phys. Rev. Lett.\/} {\bf 116} 061102 (\textit{Preprint}
  \eprint{1602.03837})

\bibitem{TheLIGOScientific:2017qsa}
Abbott B {\em et~al.\/} (LIGO Scientific Collaboration, Virgo Collaboration)
  2017 {\em Phys. Rev. Lett.\/} {\bf 119} 161101 (\textit{Preprint}
  \eprint{1710.05832})

\bibitem{GBM:2017lvd}
Abbott B {\em et~al.\/} (LIGO Scientific Collaboration, Virgo Collaboration,
  Fermi GBM, INTEGRAL, IceCube Collaboration, AstroSat Cadmium Zinc Telluride
  Imager Team, IPN Collaboration, Insight-HXMT Collaboration, ANTARES
  Collaboration, Swift Collaboration, AGILE Team, 1M2H Team, Dark Energy Camera
  GW-EM Collaboration, DES Collaboration, DLT40, GRAWITA, Fermi-LAT
  Collaboration, ATCA, ASKAP, Las Cumbres Observatory Group, {OzGrav, DWF
  (Deeper Wider Faster Program), AST3 and CAASTRO Collaborations}, VINROUGE
  Collaboration, MASTER Collaboration, J-GEM, {GROWTH, JAGWAR, CaltechNRAO,
  TTU-NRAO and NuSTAR Collaborations}, Pan-STARRS, MAXI Team, TZAC Consortium,
  KU Collaboration, Nordic Optical Telescope, ePESSTO, GROND, Texas Tech
  University, SALT Group, TOROS Collaboration, BOOTES Collaboration, MWA, CALET
  Collaboration, IKI-GW Follow-up Collaboration, H.E.S.S. Collaboration, LOFAR
  Collaboration, LWA, HAWC Collaboration, Pierre Auger Collaboration, ALMA
  Collaboration, Euro VLBI Team, Pi of Sky Collaboration, Chandra Team at
  McGill University, DFN, ATLAS Telescopes, High Time Resolution Universe
  Survey, RIMAS, RATIR, SKA South Africa/MeerKAT) 2017 {\em Astrophys. J.
  Lett.\/} {\bf 848} L12 (\textit{Preprint} \eprint{1710.05833})

\bibitem{LIGOScientific:2018mvr}
Abbott B {\em et~al.\/} (LIGO Scientific Collaboration, Virgo Collaboration)
  2019 {\em Phys. Rev. X\/} {\bf 9} 031040 (\textit{Preprint}
  \eprint{1811.12907})

\bibitem{Abbott:2020niy}
Abbott R {\em et~al.\/} (LIGO Scientific Collaboration, Virgo Collaboration)
  2021 {\em Phys. Rev. X\/} {\bf 11} 021053 (\textit{Preprint}
  \eprint{2010.14527})

\bibitem{GWTC3}
Abbott R {\em et~al.\/} (The LIGO Scientific Collaboration, the Virgo
  Collaboration and the KAGRA Collaboration) 2021 {\em arXiv e-prints\/}
  (\textit{Preprint} \eprint{2111.03606})

\bibitem{Abbott_2021}
Abbott R {\em et~al.\/} 2021 {\em The Astrophysical Journal Letters\/} {\bf
  913} L7 \urlprefix\url{https://doi.org/10.3847/2041-8213/abe949}

\bibitem{PhysRevD.103.122002}
Abbott R {\em et~al.\/} (LIGO Scientific Collaboration and Virgo Collaboration)
  2021 {\em Phys. Rev. D\/} {\bf 103}(12) 122002
  \urlprefix\url{https://link.aps.org/doi/10.1103/PhysRevD.103.122002}

\bibitem{Virgo_env_O3}
Fiori I {\em et~al.\/} 2020 {\em Galaxies\/} {\bf 8} ISSN 2075-4434
  \urlprefix\url{https://www.mdpi.com/2075-4434/8/4/82}

\bibitem{LIGO_env_O3}
Nguyen P {\em et~al.\/} 2021 {\em Classical and Quantum Gravity\/}
  \urlprefix\url{http://iopscience.iop.org/article/10.1088/1361-6382/ac011a}

\bibitem{Washimi_2021}
Washimi T {\em et~al.\/} 2021 {\em Classical and Quantum Gravity\/} {\bf 38}
  125005 \urlprefix\url{https://doi.org/10.1088/1361-6382/abf89a}

\bibitem{Punturo_2010}
Punturo M {\em et~al.\/} 2010 {\em Classical and Quantum Gravity\/} {\bf 27}
  194002 \urlprefix\url{https://doi.org/10.1088/0264-9381/27/19/194002}

\bibitem{ACERNESE2004629}
Acernese F {\em et~al.\/} 2004 {\em Astroparticle Physics\/} {\bf 20} 629--640
  ISSN 0927-6505
  \urlprefix\url{https://www.sciencedirect.com/science/article/pii/S0927650503002603}

\bibitem{ACERNESE2020102386}
Acernese F {\em et~al.\/} 2020 {\em Astroparticle Physics\/} {\bf 116} 102386
  ISSN 0927-6505
  \urlprefix\url{https://www.sciencedirect.com/science/article/pii/S0927650519301835}

\bibitem{galaxies8040085}
Allocca A {\em et~al.\/} 2020 {\em Galaxies\/} {\bf 8} ISSN 2075-4434
  \urlprefix\url{https://www.mdpi.com/2075-4434/8/4/85}

\bibitem{o3virgodetchar}
Acernese F {\em et~al.\/} 2022
  \urlprefix\url{https://arxiv.org/abs/2205.01555}

\bibitem{env_mon1}
Barone F, De~Rosa R, Eleuteri A, Milano L and Qipiani K 2002 {\em IEEE
  Transactions on Nuclear Science\/} {\bf 49} 405--410

\bibitem{vlf_website}
Romero R {\em et~al.\/} {RADIO WAVES below 22 kHz}
  \urlprefix\url{http://www.vlf.it}

\bibitem{OpenStreetMap}
{The OpenStreetMap contributors} {OpenStreetMap}
  \urlprefix\url{https://www.openstreetmap.org}

\bibitem{Peterson}
Peterson J~R 1993 {\em Open-File Report\/}
  \urlprefix\url{http://pubs.er.usgs.gov/publication/ofr93322}

\bibitem{Koley}
Koley S {\em et~al.\/} 2017 {\em SEG Technical Program Expanded Abstracts\/}
  2946--2950 \urlprefix\url{https://doi.org/10.1190/segam2017-17681951.1}

\bibitem{longuet1950theory}
Longuet-Higgins M~S 1950 {\em Philosophical Transactions of the Royal Society
  of London. Series A, Mathematical and Physical Sciences\/} {\bf 243} 1--35

\bibitem{cessaro1994sources}
Cessaro R~K 1994 {\em Bulletin of the Seismological Society of America\/} {\bf
  84} 142--148

\bibitem{AdV+}
Flaminio R 2020 {Status and plans of the Virgo gravitational wave detector}
  {\em Ground-based and Airborne Telescopes VIII\/} vol 11445 ed Marshall H~K,
  Spyromilio J and Usuda T International Society for Optics and Photonics
  (SPIE) pp 205 -- 214 \urlprefix\url{https://doi.org/10.1117/12.2565418}

\bibitem{Coughlin_2017}
Coughlin M {\em et~al.\/} 2017 {\em Classical and Quantum Gravity\/} {\bf 34}
  044004 \urlprefix\url{https://doi.org/10.1088/1361-6382/aa5a60}

\bibitem{Biscans_2018}
Biscans S {\em et~al.\/} 2018 {\em Classical and Quantum Gravity\/} {\bf 35}
  055004 \urlprefix\url{https://doi.org/10.1088/1361-6382/aaa4aa}

\bibitem{Mukund_2019}
Mukund N {\em et~al.\/} 2019 {\em Classical and Quantum Gravity\/} {\bf 36}
  085005 \urlprefix\url{https://doi.org/10.1088/1361-6382/ab0d2c}

\bibitem{USGS_PDL}
Product distribution layer git repository
  \urlprefix\url{https://github.com/usgs/pdl}

\bibitem{dms1}
Berni F {\em et~al.\/} 2012 {The Detector Monitoring System}
  \url{https://tds.virgo-gw.eu/ql/?c=9005}

\bibitem{dms2}
{F~Berni} 2020 {DMS help manual} \url{https://tds.virgo-gw.eu/ql/?c=15469}

\bibitem{logbook_rolland_EQ_mode_OK}
 2020 Virgo logbook entry validating the use of the eq mode control
  configuration to take science-quality data
  \urlprefix\url{https://logbook.virgo-gw.eu/virgo/?r=48612}

\bibitem{INGV_website_query}
 2020 Query to the public ingv website
  \urlprefix\url{http://webservices.ingv.it/fdsnws/event/1/query?starttime=2019-04-01T15\%3A00\%3A00&endtime=2020-03-27T17\%3A00\%3A00&minmag=2&maxmag=10&mindepth=-10&maxdepth=1000&minlat=27.0&maxlat=48.0&minlon=-7.0&maxlon=37.5&minversion=100&orderby=time-asc&timezone=UTC&format=text&limit=10000}

\bibitem{INGV_website}
{INGV seismic surveillance center public website}
  \urlprefix\url{http://terremoti.ingv.it}

\bibitem{nhess-15-2019-2015}
Bernardi F {\em et~al.\/} 2015 {\em Natural Hazards and Earth System
  Sciences\/} {\bf 15} 2019--2036
  \urlprefix\url{https://nhess.copernicus.org/articles/15/2019/2015/}

\bibitem{EarlyEst_website}
Early-est: Earthquake rapid location system with estimation of tsunamigenesis
  \urlprefix\url{http://early-est.rm.ingv.it/warning.html}

\bibitem{welch1967use}
Welch P 1967 {\em IEEE Transactions on audio and electroacoustics\/} {\bf 15}
  70--73

\bibitem{ROBINET2020100620}
Robinet F {\em et~al.\/} 2020 {\em SoftwareX\/} {\bf 12} 100620 ISSN 2352-7110
  \urlprefix\url{https://www.sciencedirect.com/science/article/pii/S2352711020303332}

\bibitem{Canuel:13}
Canuel B, Genin E, Vajente G and Marque J 2013 {\em Opt. Express\/} {\bf 21}
  10546--10562
  \urlprefix\url{http://www.opticsexpress.org/abstract.cfm?URI=oe-21-9-10546}

\bibitem{W_s_2021}
W{\k{a}}s M, Gouaty R and Bonnand R 2021 {\em Classical and Quantum Gravity\/}
  {\bf 38} 075020 \urlprefix\url{https://doi.org/10.1088/1361-6382/abe759}

\bibitem{Accadia_2010}
Accadia T {\em et~al.\/} 2010 {\em Classical and Quantum Gravity\/} {\bf 27}
  194011 \urlprefix\url{https://doi.org/10.1088/0264-9381/27/19/194011}

\bibitem{SBE}
Van~Heijningen J~V {\em et~al.\/} 2019 {\em Class. Quant. Grav.\/} {\bf 36} 7
  \urlprefix\url{https://iopscience.iop.org/article/10.1088/1361-6382/ab075e}

\bibitem{Valdes:2017xce}
Valdes G, O'Reilly B and Diaz M 2017 {\em Class. Quant. Grav.\/} {\bf 34}
  235009

\bibitem{Longo:2020onu}
Longo A {\em et~al.\/} 2020 {\em Class. Quant. Grav.\/} {\bf 37} 145011
  (\textit{Preprint} \eprint{2002.10529})

\bibitem{Longo:2021avq}
Longo A {\em et~al.\/} 2022 {\em Class. Quant. Grav.\/} {\bf 39} 035001
  (\textit{Preprint} \eprint{2112.06046})

\bibitem{Li_2017}
Li H, Li Z and Mo W 2017 {\em Signal Processing\/} {\bf 138} 146--158
  \urlprefix\url{https://www.sciencedirect.com/science/article/pii/S0165168417301135}

\bibitem{Cirone_mag_pay}
Cirone A {\em et~al.\/} 2018 {\em Rev. Sci.\/} {\bf 89} 114501
  \urlprefix\url{https://doi.org/10.1063/1.5045397}

\bibitem{Schumann}
Schumann W 1952 {\em Zeitschrift Naturforschung Teil A\/} {\bf 7} 149

\bibitem{StochSchumann}
Coughlin M~W {\em et~al.\/} 2018 {\em Phys. Rev. D\/} {\bf 97}(10) 102007
  \urlprefix\url{https://journals.aps.org/prd/abstract/10.1103/PhysRevD.97.102007}

\bibitem{Kowalska_Leszczynska_2017}
Kowalska-Leszczynska I {\em et~al.\/} 2017 {\em Classical and Quantum
  Gravity\/} {\bf 34} 074002
  \urlprefix\url{https://doi.org/10.1088%2F1361-6382%2Faa60eb}

\bibitem{Naticchioni_2020}
Naticchioni L {\em et~al.\/} 2020 {\em Journal of Physics: Conference Series\/}
  {\bf 1468} 012242
  \urlprefix\url{https://doi.org/10.1088/1742-6596/1468/1/012242}

\bibitem{Volland}
{Sentman, D D} 1995 {\em {Schumann Resonances}\/} (CRC Press) chap~11

\bibitem{Washimi_2021b}
Washimi T {\em et~al.\/} 2021 {\em Journal of Instrumentation\/} {\bf 16}
  P07033 \urlprefix\url{https://doi.org/10.1088/1748-0221/16/07/p07033}

\bibitem{maurin2014database}
Maurin D, Melot F and Taillet R 2014 {\em Astronomy \& Astrophysics\/} {\bf
  569} A32

\bibitem{PhysRevLett.23.184}
Beron B~L and Hofstadter R 1969 {\em Phys. Rev. Lett.\/} {\bf 23}(4) 184--186
  \urlprefix\url{https://link.aps.org/doi/10.1103/PhysRevLett.23.184}

\bibitem{amaldi1986estimate}
Amaldi E and Pizzella G 1986 {\em Il Nuovo Cimento C\/} {\bf 9} 612--620

\bibitem{GIAZOTTO1988241}
Giazotto A 1988 {\em Physics Letters A\/} {\bf 128} 241--244 ISSN 0375-9601

\bibitem{CHIANG1992603}
Chiang J, Michelson P and Price J 1992 {\em Nuclear Instruments and Methods in
  Physics Research Section A: Accelerators, Spectrometers, Detectors and
  Associated Equipment\/} {\bf 311} 603--612 ISSN 0168-9002

\bibitem{gi-1-33-2012}
Lesparre N {\em et~al.\/} 2012 {\em Geoscientific Instrumentation, Methods and
  Data Systems\/} {\bf 1} 33--42
  \urlprefix\url{https://gi.copernicus.org/articles/1/33/2012/}

\bibitem{PI_O3b}
{Puppo, P for the Virgo Collaboration} 2021 {Parametric Instability Observation
  in Advanced Virgo} second European Physical Society Conference on
  Gravitation: measuring gravity
  \urlprefix\url{https://agenda.infn.it/event/26098/contributions/132480/attachments/83185/109525/EPS_Online2021_PI.pdf}

\bibitem{PI_CohenEtAl}
Cohen D {\em et~al.\/} 2021 {Towards optomechanical parametric instabilities
  prediction in ground-based gravitational wave detectors} (\textit{Preprint}
  \eprint{2102.11070})

\bibitem{logbook52813}
 2021 Virgo logbook entry reporting the installation of the new pigtailed eom
  \urlprefix\url{https://logbook.virgo-gw.eu/virgo/?r=52813}

\bibitem{dqr}
{The LIGO Scientific Collaboration} and {The Virgo Collaboration} 2018 {Data
  Quality Report User Documentation}
  \url{https://docs.ligo.org/detchar/data-quality-report}

\bibitem{o3ligodetchar}
Davis D {\em et~al.\/} 2021 {\em Classical and Quantum Gravity\/} {\bf 38}
  135014 \urlprefix\url{https://doi.org/10.1088/1361-6382/abfd85}

\end{thebibliography}

\end{document}